%% file: Sarzi_SauronXVI.tex
\documentclass[useAMS,usenatbib,usegraphicx]{mn2e}
\usepackage{times}
\usepackage{epsfig}
\input{my_commands}

\input{figures_commands}

\title[The SAURON project - XVI]{The SAURON project XVI: On the
  Sources of Ionisation for the Gas in Elliptical and Lenticular
  Galaxies}

\author[Sarzi et al.]{Marc Sarzi,$^{1}$\thanks{E-mail:sarzi@star.herts.ac.uk} 
Joseph C. Shields,$^{2}$ Kevin Schawinski,$^{3}$ Hyunjin Jeong,$^{4}$ Kristen 
Shapiro,$^{5}$ \newauthor
Roland Bacon,$^{6}$ Martin Bureau,$^{7}$ Michele Cappellari,$^{7}$ 
Roger L.\ Davies,$^{7}$ \newauthor
P.~Tim de Zeeuw,$^{8,9}$ Eric Emsellem,$^{8}$, Jes\'us Falc\'on-Barroso,
$^{10}$ Davor Krajnovi\'c,$^{7}$ \newauthor
Harald Kuntschner,$^{8}$ Richard M.\ McDermid,$^{11}$ Reynier F.\ 
Peletier$^{12}$  \newauthor
Remco C. E. van den Bosch,$^{13}$ Glen van den Ven,$^{14}$ Sukyoung 
K. Yi,$^{4}$\\
$^{1}$Centre for Astrophysics Research, University of Hertfordshire, 
College Lane, Hatfield, Herts, AL10 9AB, UK\\
$^{2}$Physics \& Astronomy Department, Ohio University, Athens, OH 
45701 USA\\
$^{3}$Yale Center for Astronomy and Astrophysics, Yale University, P.O. 
Box 208121, New Haven, CT06520, USA\\
$^{4}$Department of Astronomy, Yonsei University, Seoul 120-749, Korea\\
$^{5}$UC Berkeley Department of Astronomy, Berkeley, CA 94720, USA\\ 
$^{6}$Centre de Recherche Astronomique de Lyon, 9~Avenue Charles 
Andr\'e, 69230 Saint Genis Laval, France\\
$^{7}$Denys Wilkinson Building, University of Oxford, Keble Road, Oxford, 
United Kingdom \\
$^{8}$European Southern Observatory, Karl-Schwarzschild-Str~2, 85748 
Garching, Germany\\ 
$^{9}$ Sterrewacht Leiden, Universiteit Leiden, Postbus 9513, 2300 RA Leiden, 
The Netherlands\\
$^{10}$Instituto de Astrofõsica de Canarias, Canarias, Via Lactea s/n, 38700 
La Laguna, Tenerife, Spain\\ 
$^{11}$Gemini Observatory, Northern Opertations Center, 670 N. AÕohoku 
Place, Hilo, HI 96720 USA\\
$^{12}$Kapteyn Astronomical Institute, Postbus 800, 9700 AV 
Groningen,The Netherlands\\
$^{13}$Department of Astronomy, University of Texas, Austin, TX 78712, 
USA\\
$^{14}$Institute for Advanced Study, Peyton Hall, Princeton, NJ 08544, 
USA }

\begin{document}

\pagerange{\pageref{firstpage}--\pageref{lastpage}} \pubyear{2008}

\maketitle
\label{firstpage}

%
\begin{abstract}

Following our study on the incidence, morphology and kinematics of the
ionised gas in early-type galaxies we now address the question of what
is powering the observed nebular emission.
To constrain the likely sources of gas excitation, we resort to a
variety of ancillary data, we draw from complementary information on
the gas kinematics, stellar populations and galactic potential from
the \sauron\ data, and use the \sauron-specific diagnostic diagram
juxtaposing the \Oiii~$ \lambda5007$/\Hb\ and
\Ni~$\lambda\lambda5197,5200$/\Hb\ line ratios.
%
We find a tight correlation 
between the stellar surface brightness and
the flux of the \Hb\ recombination line across our sample, which
points to a diffuse and old stellar source as the main contributor of
ionising photons in early-type galaxies, with post-asymptotic giant
branch (pAGB) stars being still the best candidate based on
ionising-balance arguments.
The role of AGN photoionisation is confined to the central $2''-3''$
of an handful of objects with radio or X-ray cores.
OB-stars are the dominant source of photoionisation in 10\% of the
\sauron\ sample, whereas for another 10\% the intense and
highly-ionised emission is powered by the pAGB population associated
to a recently formed stellar subcomponent.
Fast shocks are not an important source of ionisation for the diffuse
nebular emission of early-type galaxies since the required shock
velocities can hardly be attained in the potential of our sample
galaxies.
Finally, in the most massive and slowly- or non-rotating galaxies in
our sample, which can retain a massive X-ray halo, the finding of a
spatial correlation between the hot and warm phases of the
interstellar medium suggests that the interaction with the hot
interstellar medium provides an additional source of ionisation
besides old UV-bright stars.
This is also supported by a distinct pattern towards lower values of
the \OiiioHb\ ratio.
These results lead us to investigate the relative role of stellar and
AGN photoionisation in explaining the ionised-gas emission observed in
early-type galaxies by the Sloan Digital Sky Survey (SDSS).
By simulating how our sample galaxies would appear if placed at
further distance and targeted by the SDSS, we conclude that only in
very few, if any, of the SDSS galaxies which display modest values for
the equivalent width of the \Oiii\ line (less than $\sim$2.4\AA) and
LINER-like \OiiioHb\ values, the nebular emission is truly powered by
an AGN.

\end{abstract}

%
\begin{keywords}
galaxies : elliptical and lenticular -- galaxies : ISM -- galaxies : active
\end{keywords}

%
\section{Introduction}
\label{sec:intro}

If more than 30 years of investigations have now asserted the common
presence of ionised gas in early-type galaxies
\citep[e.g.][]{Cal84,Phi86,Kim89,Shi91,Gou94,Mac96,Sar06}, the sources
powering the observed nebular emission have still to be firmly
identified.
As already revealed by the spectroscopic survey of \cite{Phi86}, in
early-type galaxies the ionised-gas emission show values for several
line ratios that are generally consistent to what observed in
low-ionisation nuclear regions, or LINERs \citep{Hec80}.
Since the nature of LINER activity in itself is somehow controversial
\citep[see][for a review]{Ho08}, it is perhaps no surprise that the
excitation mechanism for the extended LINER-like emission is still not
known.
In fact, in addition to the two most touted sources for LINER
activity, a central AGN or fast shocks, the interaction with the hot
phase of the interstellar medium \citep{Spa89} and photoionisation by
old UV-bright stellar sources \citep{Bin94} have also been suggested
as mechanisms for powering the nebular emission of early-type
galaxies.
The importance of these last two ionising source was highlighted by
the narrow-band survey of \citet{Mac96} and further stressed by the
spectroscopic observations of \citet{Gou99}, albeit only in the case
of giant and dusty elliptical galaxies.

In this context, the \sauron\ integral-field survey \citep{deZ02}
allowed a more systematic spectroscopic study of the ionised gas in
nearby early- type galaxies, to an unprecedented sensitivity limit. In
\citet[][hereafter Paper~V]{Sar06} we reported an amazing variety for
the value of the \Oiii$ \lambda5007$/\Hb\ ratio both across the
\sauron\ sample and within single galaxies, which suggests that a
number of mechanisms is responsible for the gas excitation in
early-type galaxies.
If more than one source contributes to power the extended emission of
early-type galaxies, disentangling their relative importance in nearby
objects could prove critical also to understand the gas emission
observed in distant galaxies by large-scale surveys such as the Sloan
Digital Sky Survey (SDSS), where the nebular fluxes are integrated
over large physical apertures.
Yet, the limited wavelength range of the \sauron\ data
\citep[see][]{Bac01} prevents the standard emission-line diagnostic
that is routinely used to make a first assessment of the possible
ionising sources \citep[e.g., following][]{Vei87}, so that other
pieces of information are needed to tackle this problem.
Presently, we can add to our emission-line measurements not only our
knowledge of the gravitational potential of our sample galaxies
\citep[][Paper~IV]{Cap06} and of the basic properties of their stellar
populations \citep[][Paper~VI]{Kun06}, but we can also draw from a
variety of ancillary data that have been collected and analysed over
the course of the \sauron\ project.
These range from GALEX ultraviolet images \citep{Jeo09}, Spitzer
infrared data \citep{Sha09} to mm- and cm- radio observations of the
molecular and neutral gas \citep{Com07,Mor06}.
Furthermore, for a minority of our sample galaxies it is possible to
use a \sauron-specific diagnostic diagram that uses the \NioHb\ line
ratio as a gauge for the hardness of the ionising continuum.
We are therefore in a unique position to start assessing the relative
role of the various sources of ionisation that could be responsible to
the nebular emission observed in early-type galaxies.

Our investigation will proceed as follows. In
\S\ref{sec:SAURONdiagnostic} we will introduce the potential of the
\NioHb\ diagnostic diagram and derive more robust \Ni\ measurements in
order to maximise its use across our sample. In \S\ref{sec:Ionisation}
we will review the importance of AGN activity, star formation, diffuse
old stellar sources, fast shocks and of the interaction with the hot
phase of the interstellar medium in powering the nebular emission
observed across the \sauron\ sample. In \S \ref{sec:SDSS} we will
explore the implications of the lessons learnt in \S
\ref{sec:Ionisation} for interpreting the ionised-gas emission
observed in distant early-type galaxies that are targeted by
large-scale spectroscopic survey such as the SDSS, and finally draw
our conclusions in \S \ref{sec:conclusions}.

\section{SAURON emission-line diagnostics}
\label{sec:SAURONdiagnostic}

Emission-line diagnostic diagrams have had a long history of success
in constraining the different sources of ionisation for the nebular
emission observed in extra-galactic objects.
Initially introduced by \citet[][hereafter BPT]{Bal81} to separate
star-forming objects and active galactic nuclei (AGN), these diagrams
juxtapose various emission-line ratios to provide a basis for
classifying the gas emission into categories that, to varying degrees
of confidence, can be identified with specific excitation sources.
This is achieved by comparing the position of the data in the
diagnostic diagrams with the predictions of sophisticated models for
gas that is photoionised by a central AGN or by OB-stars
\citep[\eg][]{Fer98,Kew01} or that is excited by shocks
\citep[\eg][]{Dop95,Dop96}.
Today, the most widely adopted kind of BPT-diagrams are those
introduced by \citet{Vei87}, who designed a diagnostic analysis that
is insensitive to reddening and that is based on the ratios of strong
emission lines rather close in wavelength, such as \Hb~$\lambda4861$
and \Oiii~$\lambda\lambda4959,5007$ or \Oi~$\lambda6300$,
\Nii~$\lambda\lambda6548,6583$, \Ha~$\lambda6563$, and
\Sii~$\lambda\lambda6716,6731$.

Among the emission lines that usually feature in these diagrams only
the \Oiii\ and \Hb\ lines fall in the wavelength range of the
\sauron\ spectra, which was intentionally restricted around the
\Hb\ and Mg$b$ absorption-line features in order to measure the
stellar kinematics and the properties of the stellar populations
within the largest possible field of view.
A traditional diagnostic analysis in our sample galaxies is therefore
only possible when the \sauron\ data are complemented by
integral-field observations in the \Ha+\Nii\ spectral region, as shown
by \citet{Maz06} for the Sa galaxy NGC~7742.

\subsection{The \Ni\ lines and the \Ni/\Hb\ diagnostic diagram}
\label{subsec:NI}

If standard BPT-diagrams cannot be drawn with \sauron\ data alone, the
emission from the weak \Ni~$\lambda\lambda5197,5200$ doublet allows
for a diagnostic analysis that is rather similar to that normally
carried out through the BPT-diagram based on the \OioHa\ line ratio.
Indeed, similarly to the \Oi\ emission also the \Ni\ lines arise in
partially ionised regions, which are extended in gaseous nebulae
photoionised by a spectrum containing a large fraction of high-energy
photons, but are almost absent in \Hii-regions.
The \NioHb\ ratio can therefore be used to gauge the hardness of the
ionising continuum, making the \NioHb\ vs. \OiiioHb\ diagnostic
diagram in principle as useful as the \OioHa\ vs. \OiiioHb\ diagram to
separate photoionisation by OB-stars from other excitation sources.

\placefigNioHbone 

This is demonstrated by Fig.~\ref{fig:NIone}, which shows that the
predictions of the MAPPING-III models in the event of a starburst, of
gas photoionised by a central AGN and of gas excited by shocks
\citep[from][respectively]{Dop00,Gro04b,All08} are as well separated
in the \sauron\ \NioHb\ diagnostic as they are in the standard
\OioHa\ BPT diagram.
In Fig.~\ref{fig:NIone} we also show the distribution of the
emission-line ratios measured with the method of Paper~V in the
central regions of 284 galaxies observed by the Sloan Digital Sky
Survey (SDSS). This sample includes all galaxies with SDSS spectra
(from the DR6 release), not only early-type galaxies, with redshifts
between $0.02<z<0.05$, apparent $r$-band magnitude brighter than
$r<16$ and, most important, with detected \Ni\ emission.
If we adopt the conservative theoretical limit of \citet{Kew01} to
identify star-bursting SDSS galaxies in the \OioHa\ BPT diagram and
further separate Seyfert nuclei from objects showing LINER-like
central emission using the empirical line drawn by \citet{Kew06},
Fig.~\ref{fig:NIone} illustrates how these three kinds of objects are
clearly separated also in the \sauron\ \NioHb\ diagnostic diagram.

The position of the SDSS galaxies displaying nuclear star formation
and LINER-like emission in Fig.~\ref{fig:NIone} deserves further
attention.
Starting with the star-forming galaxies, the fact that the
MAPPINGS-III starburst models systematically underpredict the values
of the \NioHb\ ratio suggests that, similarly to case of the \Oi\ line
\citep{Sta96,Dop00}, also the \Ni\ fluxes can be significantly
enhanced by the mechanical energy released by supernovae and stellar
winds, which are naturally expected in starbursts as a result of
stellar evolution.
As regards the location of the SDSS galaxies exhibiting central
LINER-like emission, we note that these objects are more clearly
separated from the Seyfert nuclei in the \NioHb\ diagnostic diagram
than they are in the \OioHa\ diagram. Furthermore, most of the SDSS
LINER nuclei appear to follow a tight sequence consistent with
photoionisation by a central AGN with increasingly harder ionising
radiation fields, in agreement with the findings of \citet{Kew06}
based on the \OioHa\ diagnostic.
These results further stress the usefulness of the \Ni\ lines as a
gauge for the hardness of the ionising continuum.

Unfortunately, the $2^{\rm nd}$ and $3^{\rm rd}$ energy levels of
neutral Nitrogen have a much smaller critical de-excitation density
($7\times10^2$ and $2.2\times10^3$ \pccm) than the $4^{\rm th}$ level
of neutral Oxygen ($1.5\times10^6$ \pccm), so that the \Ni\ lines are
generally much fainter than the \Oi\ line owing to collisional
de-excitation. The \sauron\ \NioHb\ diagnostic diagram is therefore
applicable only in the few early-type galaxies where nebular emission
is particularly intense.
To complicate matters further, as discussed in Paper~V the detection
of the \Ni\ lines is hampered by our limited ability to match the
underlying continuum in the spectral region around the Mg$b$
absorption features.

\subsection{Improving the \Ni\ measurements with the MILES spectral library}
\label{subsec:Miles}

In \citet[][Paper~X]{Cap07} we re-extracted the stellar kinematics for
our sample galaxies using for each galaxy a different subset of
stellar templates from the MILES library of \citet{San06}.
The MILES templates provide a much better fit to the emission-free
spectral regions of our data, and thus their use delivers a more
reliable stellar kinematics.

In an attempt to further improve the extent and quality of our
\Ni\ measurements, we re-extracted also the nebular emission fluxes
and kinematics using the MILES stellar library.
In the case of emission-line measurements, however, it is crucial to
provide an underlying fit to the stellar continuum that is as
physically motivated as possible in order to avoid spurious
results. For this reason, rather than using a subset of stellar
spectra from the MILES library for each galaxy as done in Paper~X, we
used a template library consisting of simple stellar population models
from \citet[][based on MILES stellar spectra, see also
  \citealt{Vaz07}]{Vaz09} to which we added a number of empirical
templates obtained by matching real \sauron\ spectra, most of the
times devoid of emission.

Specifically, based on the maps of Paper~V for the ionised-gas
emission, we extracted a number of high-S/N \sauron\ spectra from
circular apertures in galactic regions either without significant
emission or with emission lines that were clearly distinguishable from
the underlying absorption spectrum, which we fitted over the {\it
  entire} wavelength range with the MILES stars (excluding stellar
templates with exceedingly low values of the H$\beta$ absorption-line
strength).
The optimal combination of the stellar spectra that best matched each
of our emission-free aperture spectra constitute each of the empirical
templates in our new template library.
Care was taken to extract \sauron\ spectra covering the distribution
of the absorption-line strengths observed in Paper~VI, which cannot be
covered in detail by the distribution of line-strengths values
measured in the MILES stellar library since this consists mostly of
spectra for nearby stars.
In this respect, our empirical templates approximate as closely as
possible the spectra of real early-type galaxies, were they unaffected
by kinematic broadening.
Fig.~\ref{fig:MILESone} illustrates the poor coverage that the MILES
stellar library provides for the absorption-line strengths observed in
early-type galaxies, in particular in the plane defined by the Mg$b$
and Fe5015 indices, which the adopted array of single stellar
population models and empirical templates can match in a more physical
manner.

\placefigMILESone

Compared to the fit of Paper~V, the use of such a mixed library of
model and empirical templates based on MILES stellar spectra leads to
a significant increase in the quality of the fit to the
\sauron\ spectra in the Mg$b$ region (corresponding to a 30\% decrease
in the RMS of the fit residuals), which adds confidence in the
\Ni\ measurements.
In general the \Hb\ and \Oiii\ measurements obtained with the new set
of templates do not differ significantly from our previous
measurements, which reflect the modest 10\% decrease in the RMS of the
fit residual when the entire \sauron\ wavelength range is considered.
Therefore, for the sake of clarity hereafter we will quote and plot
the \Hb\ and \Oiii\ measurements from Paper~V, and use the newly
extracted emission-line measurements only when discussing the
\NioHb\ diagnostic diagrams or when comparing maps for the \Ni\ and
\Hb\ emission.

\placefigAGNone 
\placefigAGNtwo 
\placefigAGNthree

\section{Ionisation Sources}
\label{sec:Ionisation}

Except for very low and high values of the \OiiioHb\ ratio that most
likely correspond to \Hii-regions and Seyfert nuclei, respectively,
the diagnostic diagrams of Fig.~\ref{fig:NIone} illustrate vividly how
without other line ratios than \OiiioHb\ it is impossible to
immediately separate star formation from AGN activity.
Besides, a number of ionising mechanisms in addition to AGN activity,
such as fast shocks \citep{Dop95}, photoionisation by old stars
\citep{Bin94} and the interaction with the hot, X-ray emitting gas
\citep{Spa89}, have been suggested to explain the emission in the
log(\OiiioHb)= 0 -- 0.5 range, which is typical of LINER-like emission
\citep[e.g.,][]{Ho97} and that includes 75\% of \OiiioHb\ values
observed across the \sauron\ sample.
In fact, very low \OiiioHb\ values are observed also in the absence of
star formation, for instance in the X-ray filaments around the
brightest galaxies of clusters \citep[e.g.,][]{Hat06,Sab00}, whereas
quite large \OiiioHb\ ratios can be observed also in \Hii-regions if
the metallicity is extremely low.

For this reason in this section we will use a variety of ancillary
data, ranging from radio to X-ray wavelengths, and consider
complementary information on the gas morphology and kinematics from
our \sauron\ data to constrain the most likely sources of gas
excitation in different kind of galaxies, using, when possible, the
\NioHb\ diagnostic to corroborate our findings.

\subsection{Nuclear Activity}
\label{subsec:AGN}

Many of our sample galaxies host AGNs. Well-known cases include
NGC~4486 and NGC~4374 for their FR I radio jets \citep[\eg][]
{Owe89,Lai87}, NGC~4278 for hosting a well-studied LINER nucleus
\citep[\eg][]{Gou94,Gir05} and NGC~4552 for an ultraviolet flare that
could have resulted from the tidal stripping of a star passing close
to the central black hole \citep{Ren95,Cap99}

Given that unresolved radio-continuum cores are excellent signposts
for black-hole accretion, in this section we focus on the galaxies in
our sample that exhibit such a feature. In the area covered by the VLA
FIRST survey \citep{Bec95} there are 15 galaxies with 1.4~Ghz compact
cores. Figure~\ref{fig:AGNone} compares the \sauron\ \Hb\ emission
with the distribution of the 1.4~Ghz radio continuum, which in several
cases extends beyond the nucleus.
In particular, a comparison with the beam of the FIRST images reveals
that the nuclear emission is resolved in NGC~2974, NGC~3414, NGC~4546,
NGC~5813 and NGC~5846, while in NGC~3489 the radio core is only
marginally extended.

Among the \sauron\ galaxies that have not been observed by the FIRST
survey, the presence of a radio core has been excluded in NGC~474 and
NGC~821 by \citet{Wro91} and in NGC~7457 and NGC~7332 by
\citet{Mor06}. The remaining galaxies, NGC~524 and NGC~1023, do host a
radio core according to \citet{Fil04} and Morganti et al.,
respectively.
In what follows we will use the radial profiles for the values of the
\Hb\ flux and the \OiiioHb\ ratio to understand the extent to which
AGN can power the nebular emission in our sample galaxies. In
particular, it will be shown that extremely shallow \Hb\ and
\OiiioHb\ are inconsistent with AGN photoionisation.  This is the case
of NGC~524, whereas for NGC~1023 the absence of central \Hb\ emission
precludes any further analysis.
Conversely, among the objects with undetected radio-continumm cores in
the FIRST images, NGC~4477 display a strong and unresolved central
\Hb\ peak and a very steep \OiiioHb\ gradient, suggesting that AGN
photoionisation is important toward the centre of this galaxy. The
presence of a powerful AGN in NGC~4477 is signalled by a strong X-ray
nucleus \citep[first detect by the {\it Einstein} observatory
  by][]{Fab92}, and its radio-quiet nature is consistent with its
Seyfert classification \citep[Ho et al. 1997;][]{Ho08}.

If we assume a uniform gas density and constant filling factor for gas
clouds residing on a plane, in ionisation equilibrium the surface
brightness of the \Hb\ recombination line should follow the strength
of the radiation field, therefore radially decreasing as $r^{-2}$ in
the case of a central source.
The strength with which the AGN radiation reaches the gas clouds,
usually quantified by the ionisation parameter $q$ that compares the
ionising-photon and electron density at the face of the irradiated
cloud, also determines the ionisation state of the gas clouds.
For our purposes the \OiiioHb\ ratio is an excellent diagnostic for
the ionisation state of the gas, not only since this ratio involves
emission from a highly-ionised species \citep[\eg][]{Fer83,Ho93} but
also because the \OiiioHb\ ratio scales almost linearly with $q$ for
the range of values observed in the \sauron\ nuclei (i.e., for
\OiiioHb $\le 5$).
Therefore, also the values of the \OiiioHb\ ratio should follow a
$r^{-2}$ profile under the previous assumptions.

On the other hand, if the gas density decreases with radius, the
\Hb\ surface brightness will drop at an even faster gradient than
inverse-square, whereas the \OiiioHb\ ratio will display a shallower
radial profile as the ionisation parameter $q$ decreases more gently
with radius.
Furthermore, if the gas is also vertically extended rather than just
being distributed in a planar configuration, projection effects will
lead to shallower \Hb\ surface brightness profiles. In this case the
\OiiioHb\ values will follow the same trend as the \Hb\ emission and
also show a gentler radial decline, as clouds at different distances
from the centre and hence subject to different ionising photon
densities are observed along the same line of sight.
Conversely, if the gas clouds lie in a plane, the AGN radiation could
be absorbed by intervening material before reaching the gas clouds at
any given radius, leading to \Hb\ and \OiiioHb\ profiles that are
steeper than an $r^{-2}$ law.
Finally, radial gradients in the filling factor of the gas clouds will
only affect the \Hb\ surface-brightness profiles, in the same
direction as variations in the gas density would do.

Figure~\ref{fig:AGNtwo} plots the radial profile of the \Hb\ emission
in the \sauron\ galaxies with 1.4~Ghz radio cores, together with the
predicted $r^{-2}$ radial decrement for AGN photo-ionisation and under
the special conditions of constant gas density and filling factor,
accounting for atmospheric blurring. The \Hb\ data-points in
Fig.~\ref{fig:AGNtwo} are colour-coded according to the value of the
\OiiioHb\ ratio, in order to convey also the radial variation of this
quantity.
Figure~\ref{fig:AGNtwo} shows that the surface brightness profile of
the \Hb\ emission is in keeping with our simplest expectations for AGN
photo-ionisation only in the very central regions of a few objects
(\eg\ NGC~3414, NGC~4278, NGC~4546) whereas in general the
\Hb\ profile is shallower than an inverse-square law, in particular
beyond the central $\sim3''$.
The behaviour of the \OiiioHb\ ratio, which does not appear to
systematically decrease with radius, further rules out a central AGN
as the source of ionisation for the gas emission beyond the innermost
regions of our sample galaxies.
In fact, that the \OiiioHb\ ratio tends to display rather constant
values in the innermost regions (within 20\% of the average within
3$''$, see also Fig.~4b of Paper~V) suggests that if an AGN is
responsible for the gas excitation where the \Hb\ surface-brightness
profile is consistent with a $r^{-2}$ law, then the assumptions of our
simplest scenario cannot hold.
In this respect, we note that if the central gas of our sample
galaxies has settled on a steady disk, this should display a radially
decreasing profile for the column density \citep{Kaw07}, which is
proportional to the gas density and the filling factor of the gas
clouds.
The presence of a negative gradient for the gas density could address
the need for shallow \OiiioHb\ profiles in the nuclear regions of
galaxies with \Hb\ gradients that are consistent with the
inverse-squared model, even though a radially decreasing gas density
would induce even steeper \Hb\ gradients. Indeed, such a change could
not be appreciated by our observations as those central \Hb\ profiles
are in fact unresolved.
On the other hand, if we were to advocate the role of AGNs also where
the \Hb\ surface-brightness profile is resolved and is slightly
shallower than an $r^{-2}$ law (\eg\ NGC~2768, NGC~2974, NGC~4486) 
then projection effects would have to be invoked in addition to a radial
decrement of the gas density.
Considering vertically extended structures may also help to explain
some of the sharpest peaks for the value of the gas velocity
dispersions that are observed in our sample (\eg\ NGC~3414, NGC~4278,
NGC~5198).

As a final item of this section, in Fig.~\ref{fig:AGNthree} we draw
the \NioHb\ vs. \OiiioHb\ diagnostic diagram for the ionised-gas
emission observed in the central 3$''$ of the \sauron\ galaxies with
1.4~Ghz radio-continuum cores and \Ni\ emission.
The emission in the central regions of these objects is generally
consistent with both photoionisation from a central AGN with a very
low ionisation parameter and with excitation from fast shocks. If we
disregard the latter possibility for the moment (see
\S\ref{subsec:Shocks}), we note that only models including gas of
super-solar metallicity and a self-consistent treatment of dust
\citep{Dop02,Gro04b} provide a plausible description for our data.
In particular, dust-free models with gas of solar metallicity
(Fig.~\ref{fig:AGNthree}, left) can match just a fraction of our data
(\eg\ the very center of NGC~4486) and only with the hardest ionising
radiation fields (\ie\ $f_\nu\propto\nu^{-1.2}$).
Super-solar values for the gas metallicity are required also in the
case of dusty, radiation pressure-dominated photoionization
(Fig.~\ref{fig:AGNthree}, right), since also these models would have
to invoke the hardest ionising continuum in order to match our data in
the case of gas of solar metallicity (not shown for the sake of
clarity).
The very nuclear regions of the objects in Fig.~\ref{fig:AGNthree}
(within 1$''$, smallest symbols) appear consistent with dusty
super-solar AGN models while adopting similarly low values of the
ionisation parameter and a realistic range for the hardness of the AGN
radiation field (from $f_\nu\propto\nu^{-1.7}$ to $\nu^{-1.2}$), in
keeping with the behaviour of the SDSS LINER nuclei
(\S\ref{subsec:NI}).
Figure~\ref{fig:AGNthree} shows that significant radial variations in
the average values of the \NioHb\ and \OiiioHb\ ratios are found only
in NGC~2768, NGC~4278 and NGC~4486. As the average values of the line
ratios do not run parallel to the dusty AGN models, such gradients are
inconsistent with a simple radial change in the mean value of the
ionising parameter (due to either a different mean distance or gas
density) or of the gas density, suggesting instead a variation in the
hardness of the AGN radiation field. This seems rather unlikely,
however, in particular for NGC~4486 where the radial increase of the
\NioHb\ ratio would imply a hardening of the AGN continuum hitting
clouds at increasing distance from the centre. Therefore, excluding
the possibility of special gas and dust geometries, we consider it
more likely that other processes than AGN photoionisation are
responsible for the gas excitation beyond the very centre of these
objects. In fact, for NGC~4486 the results of \citet{Sab03} and
\citet{Dop97} suggest a radial transition from nuclear AGN
photoionisation to circumnuclear gas excited by fast shocks.\\


To conclude, the presence of a 1.4~Ghz radio-continuum core suggests
that a central AGN could be responsible for powering the ionised-gas
emission in the very centre of 30\% of our \sauron\ sample galaxies,
at most. For AGN photoionisation to be consistent with the observed
radial profiles of the \OiiioHb\ line ratio within the central 3$''$,
the central gas distribution has to have a radially decreasing
density, and in the case of objects with resolved \Hb\ emission
profiles, also a vertically extended structure. Even so, the presence
in few objects of considerable gradients in the value of \NioHb\ ratio
suggests that other mechanisms than photoionisation by a central AGN
must contribute to power the nebular emission outside the very centre.

\subsection{\Hii-regions}
\label{subsec:Ostars}

A few early-type galaxies in the \sauron\ sample show clear evidence
of on-going star-formation.
In particular, the values of the \OiiioHb\ ratio in NGC~3032,
NGC~4526, NGC~4459 plunge to levels that are well below the
lower-limit of the \OiiioHb\ ratio that is observed also in LINER-like
emission regions and Seyfert nuclei, which is around \OiiioHb$\sim0.3$
or log(\OiiioHb)$=-0.5$ \citep[see, \eg][]{Kew06}. As noticed already
in Paper~V, these objects are also characterised by extremely regular
dust morphologies and gas kinematics, in particular that of the
\Hb\ line, suggesting a relaxed gas distribution and
dynamics. Similarly relaxed structures are observed also in NGC~524
and NGC~5838, which too show low \OiiioHb\ values although never down
to the low level that is observed in the previous galaxies.
That star formation may be occurring in NGC~3032, NGC~4459, and
NGC~4526 is supported also by the finding of massive disks of
molecular gas \citep{Com07,You07} and the detection of emission from
polycyclic aromatic hydrocarbons \citep[PAHs][]{Sha09}.

Yet, even in these relatively simple objects we observe strong
gradients in the value of the \OiiioHb\ ratio, which underscore the
presence of other sources of ionisation than OB-stars.
In this respect it is useful to observe in Fig.~\ref{fig:SFone} the
behaviour in the \NioHb\ diagnostic diagram of the \sauron\ data for
NGC~3032 and NGC~4526, the only two star-forming early-type galaxies
with detected \Ni\ emission. For comparison, in Fig.~\ref{fig:SFone}
we also show the same \sauron\ measurements for the Sa galaxy NGC~4314
\citep[from][]{Fal06}, which hosts a central star-forming ring and
where the nebular emission is stronger, in terms of equivalent width
of the lines, than in NGC~3032 and NGC~4526.
Figure~\ref{fig:SFone} confirms dramatically the presence of star
formation in NGC~3032 and NGC~4526 as it shows that the
\sauron\ measurements in these two objects display the same
characteristic diagonal trend in the \NioHb\ vs. \OiiioHb\ diagram
that is observed also in SDSS galaxies with star-bursting nuclei
(Fig.~\ref{fig:NIone}) and in the ring of NGC~4314.
Furthermore, Fig.~\ref{fig:SFone} confirms the impression conveyed by
the maps for the equivalent width of the \Hb\ emission (Fig.~4b of
Paper~V) that, as in the case of NGC~4314, star formation in NGC~3032
and NGC~4526 occurs primarly in a ring. Indeed, in
Fig.~\ref{fig:SFone} the \sauron\ data reveal a striking similarity
between the line-ratio gradients observed in NGC~3032, NGC~4526 and
the Sa NGC~4314, whereby in all three galaxies the emission-line
properties move closer to the predictions of the dusty AGN or shock
models as we consider regions away from the star-forming ring {\it
  both\/} towards the centre and the outer regions of the galaxy.
In fact, in NGC3032 the regions outside the ring seem to approach the
AGN and shock grids from a different direction than the points inside
the ring, which would suggest either that toward the centre AGN
photoionisation is more important than shock excitation, or that
shocks may occur under different conditions at the opposite edges of
the ring. A similar trend was observed also by \citet{Maz06} in the Sa
galaxy NGC~7742 using standard BPT diagrams.

\placefigSFone

In Fig.~\ref{fig:SFone} we also overplot on the \sauron\ maps the
contour corresponding to the 1.4~Ghz radio-continuum emission measured
by the FIRST survey. Consistent with synchrotron radiation from
relativistic electrons and free-free emission from \Hii\ regions, the
radio continuum is resolved and matches well the distribution of the
nebular emission.
Similarly to the case of normal disk galaxies \citep{Ken83}, the
1.4~Ghz emission in NGC~3032 and NGC~4526 is dominated by the
non-thermal synchrotron component, given that the thermal free-free
emission that would correspond to our \Hb\ fluxes falls short to
explain the observed flux density values.
In fact, following the prescription of \citet{Con92} and \citet{Cap86}
to convert the \Hb\ emission into free-free emission, the observed
1.4~Ghz flux densities appear remarkably consistent with our
\sauron\ emission-line measurements (accounting for beam smearing) if
we adopt a 10\% fraction for the thermal component to the total
1.4~Ghz fluxes, as found on average in disk galaxies \citep{Ken83}.
That also in early-type galaxies the 1.4~Ghz continuum traces mostly
the synchrotron emission may explain why in Fig.~\ref{fig:SFone} the
radio contours do not appear to follow in detail the \Hb\ emission
observed in NGC~3032 and in the Sa NGC~4314, in particular where the
\Hb\ fluxes peak. Judging the agreement between radio and nebular
emission is more difficult in NGC~4526, due to the higher inclination
of its gaseous disk. It is instructive to note also the different
impact of reddening by dust in NGC~3032 and NGC~4526 due to
inclination effects, which explains why the latter galaxy shows
brighter radio continuum flux densities, which are unaffected by dust,
despite displaying fainter \Hb\ emission.

No diffuse radio emission is detected in the rings of NGC~4459 and
NGC~524, due to the fact that the nebular emission in these objects is
significantly fainter than that observed in NGC~3032 and NGC~4526, and
the corresponding radio emission has escaped detection in the FIRST
images.
Given the typical detection threshold of 0.45 mJy/beam in the FIRST
survey, we expect to detect radio emission associated to
star-formation in the FIRST images only if the average surface
brightness of the \Hb\ line within 99.73\% of the FIRST beam
(FWHM=5\farcs4) exceeds ${\rm \sim0.9\times10^{-16}
  erg\,cm^{-2}s^{-1}arcsec^{-2}}$, including also the contribution of
synchrotron emission.
This level of surface brightness for the \Hb\ line (conveniently shown
in Fig.~4a of Paper~V in light-green colours) is never achieved in the
ring regions of NGC~524 and NGC~4459.
On the other hand, the \Hb\ and \Oiii\ flux values in NGC~5838 are
comparable to what is measured in NGC~4526, suggesting that the
unresolved 1.4~Ghz emission shown in Fig.~\ref{fig:AGNone} could arise
from the compact gaseous disk of NGC~5838 as well as from a central
AGN, in agreement also with the fact that this object sits on the
far-IR to radio-continuum emission relation \citep{Com07}.

Could photoionisation by OB-stars be responsible also for the nebular
emission that is observed in galaxies with less relaxed dust and gas
structures or log(\OiiioHb) values greater than $-0.5$?
Star formation is unlikely to be the source of excitation for clouds
with log(\OiiioHb)$\, >0.5$, corresponding to starburst models with
low metallicity and large $q$ values (Fig.~\ref{fig:NIone}). Star
formation episodes characterised by subsolar metallicity values are
indeed extremely rare (because gas is quickly enriched during such
events), whereas the ionisation parameter of extragalactic
\Hii-regions is typically found to range only between $10^7<q<10^8$
\citep{Dop00}.
For more intermediate values of the \OiiioHb\ ratio, the possibility
that OB-stars could be powering the observed emission is particularly
compelling when there is evidence for recent star formation, for
instance from the analysis of optical spectra (Paper~VI and VIII) or
of near-UV colours \citep[\eg][]{Jeo07}, or in the presence of massive
reservoirs of molecular gas \citep{Com07}.
In this respect, cases like NGC~3489 and NGC~3156 are particularly
intriguing (but see also \S\ref{subsec:HIERs}). Both of the previous
conditions are met in these objects, where the gas kinematics is also
fairly relaxed with small values of the gas velocity dispersion
\Sg. Yet, despite these similarities with the star-forming objects,
the \OiiioHb\ ratio in NGC~3489 and NGC~3156 is consistently $> 1$,
peaking to remarkably high values ($\sim$10) in places well outside
the nucleus, which rules out \Hii\ regions as the main source of
nebular emission.

For these and other objects with evidence of recent star formation or
reservoirs of molecular gas we can use the VLA FIRST radio-continuum
data to further test whether the observed nebular fluxes are {\it
  entirely\/} associated to star formation activity.
For instance, if the nebular emission observed in NGC~3489 was powered
by OB-stars, following the previous prescription to derive 1.4~Ghz
fluxes from the observed \Hb\ emission, we should see a well resolved
radio structure, as in the case of the star-forming galaxy NGC~4526
that indeed displays similar values for the \Hb\ fluxes.
This is not observed, since NGC~3489 presents a rather weak and only
marginally resolved core (\S\ref{subsec:AGN})
Similarly, in NGC~3156 the central \Hb\ emission is sufficiently
intense that if it were associated to star formation it would lead to
a detectable radio core, but this is not observed.
The non detection of 1.4~Ghz radio-continuum features excludes the
possibility that nebular emission is powered solely by OB-stars also
in the nuclei of NGC~2685, NGC~4150 and NGC~7332, which have young
stellar populations, relatively bright central \Hb\ emission and,
except for NGC~7332, also molecular gas.
Among the objects with old stellar ages, as estimated from the
\sauron\ absorption-line measurements, it is worth considering the
cases of NGC~2768, which has a robust single-dish CO detection
\citep{Com07}, and of NGC~2974, where the near-UV colours suggest the
presence of young stars in the centre and in an outer ring
\citep{Jeo07}. Although both galaxies display strong radio-continuum
cores that are undoubtably dominated by AGN activity, the observed
\Hb\ fluxes are sufficiently strong and extended (exceeding the ${\rm
  \sim0.9\times10^{-16} erg\,cm^{-2}s^{-1}arcsec^{-2}}$ threshold,
Fig.~\ref{fig:AGNtwo}) that if they were entirely powered by OB-star
the corresponing free-free and synchrothron emission from star-forming
regions would be capable of inducing noticeable and extended features
in the radio-continuum maps.
That in NGC~2768 the radio core is unresolved therefore suggests that
star-formation cannot be principal the source of ionization in the
central regions. Conversely, the fact that the radial profile for the
1.4Ghz continuum in NGC~2974 is resolved does not necessarly imply
that the \Hb\ emission is powered by OB-stars. In fact, this
possibility will be ruled out in \S\ref{subsec:Shocks} by means of the
\sauron\ \NioHb\ diagnostic, suggesting that the extended nature of
the radio core of NGC~2974 may be due to the presence of an unresolved
jet.
Nonwithstanding the lack of evidence for nuclear star formation, we
note that NGC~2768 and NGC~2974 are among the brightest IRAS
far-infrared sources in our sample (together with NGC~524, NGC~3032,
NGC~4459, NGC~4526 and NGC~5838) suggesting that star-formation occurs
primarily outside the centre in these two galaxies.\\


To conclude, extremely low values of the \OiiioHb\ ratio leave no
doubt that OB-stars are powering most of the nebular emission observed
in the 6\% of the \sauron\ sample galaxies (3/48), with regular dust
morphologies and relaxed gas kinematics strongly suggesting that star
formation could be responsible for the gas ionisation in up to 10\% of
the sample (5/48).
The presence of young stellar populations, molecular gas, and the
detection of PAHs features and strong FIR fluxes further support the
case for on-going star formation in these objects.
For two of them, NGC~3032 and NGC~4526, the \sauron\ \NioHb-
diagnostic diagram reveal a dramatic similarity with the star
formation activity in the central rings of the early-type disk
galaxies, showing also the transition to regions outside the ring
where other excitation mechanisms appear to dominate, such as AGN
photoionisation towards the centre.
The detection in the FIRST data for NGC~3032, NGC~4526 and NGC~5838 of
radio-continuum emission consistent with synchrotron radiation from
relativistic electrons and free-free emission from \Hii-region further
supports the case for on-going star-formation in these objects,
whereas the non-detection of extended 1.4 Ghz emission in NGC~524 and
NGC~4459 can be ascribed to a lower star-formation activity and the
limited sensitivity of the FIRST survey.
The FIRST data also suggest that in other objects with young stellar
populations and molecular reservoirs only a fraction of the
\Hb\ fluxes can be attributed to \Hii-regions.

\placefigDiffuseone
\placefigDiffusetwo

\subsection{Diffuse Evolved Stellar Sources}
\label{subsec:OldStars}

Apart from very young stars, several kinds of evolved stars or stellar
remnants emit significant amounts of ionising photons, such as
post-asymptotic giant branch stars (pAGB), extreme horizontal-branch
stars and X-ray binaries. Although none of these objects produce an
ionising continuum as powerful as that of a single O-star, their
importance as photoionisation sources lies in their large numbers and
potential ubiquity, in particular since early-type galaxies are made
mostly of old stellar populations.
In fact, the significance of pAGB stars as ionising candidates was
first demonstrated by \citet{Bin94}, who showed that when taken
together such objects produce a sufficient Lyman continuum to account
for the \Ha\ luminosity of early-type galaxies.
Shortly afterwards, \citet{Mac96} reported the finding of a
correlation 
%
\citep[already hinted at by][]{Phi86}
%
between the \Ha+\Nii\ luminosity of early-type galaxies and their
integrated B-band luminosity within the typical extent of the nebular
emission, thus adding to the hypothesis that the principal sources of
ionising photons are to be found within the bulk of the stellar
population.

\subsubsection{The Connection Between Nebular and Stellar Emission}

Following \citeauthor{Mac96} we computed for each of our sample
galaxies the average distance from the center of the detected
emission-line regions, and integrated the flux of both the stellar
continuum and of the \Hb\ recombination line observed over all the
\sauron\ spectra within such a radius.
After rescaling for galactic distance, the left panel of
Fig.~\ref{fig:Diffuseone} shows a nearly linear correlation between
the luminosity of the \Hb\ line and that of the stellar continuum
within the average extent of the nebular emission, consistent with the
narrow-band results of \citeauthor{Mac96}

Yet, the true strength of the link between the nebular and stellar
emission in early-type galaxies is dramatically revealed when the
total \Hb\ luminosity is compared with the luminosity of the stellar
continuum if this is integrated only {\it where\/} the nebular
emission is detected.
As the central panel of Fig.~\ref{fig:Diffuseone} illustrates, in this
case the correlation between the values of the \Hb\ luminosity $L_{\rm
  H\beta}$ and continuum luminosity $L_{\rm cont}$ is much
tighter\footnote{With the benefit of hindsight a tight correlation can
  also be observed in the measurements of \citeauthor{Mac96} if only
  the objects with diffuse and extended emission (dubbed ``DE'') are
  considered.}.
Such a relation is nearly, although not quite, linear (with a
power-law slope $a=1.30\pm0.08$), meaning that the equivalent width of
the {\it integrated\/} \Hb\ emission is almost constant across our
sample.
The tightness of $L_{\rm cont}$--$L_{\rm H\beta}$ relation (with a
0.15 dex scatter) is quite remarkable considering that in addition to
a diffuse stellar sub-population other sources of ionisation may
contribute to the total flux of the \Hb\ recombination line.
In fact, the only obvious outlier in Fig.~\ref{fig:Diffuseone} is
NGC~3032, the galaxy displaying by far the most intense star-forming
activity in our sample.
The possible contribution of a central AGN does not affect much the
$L_{\rm cont}$--$L_{\rm H\beta}$ relation, as its power-law slope and
scatter do not change significantly ($a=1.20\pm0.05$, 0.13 dex
scatter) when the central $3''$ are excluded (\S\ref{subsec:AGN}).
The tightness of this correlation is not a distance artifact either,
since it persists also when the flux of the \Hb\ emission and of the
stellar continuum are compared (with $a=1.31\pm0.08$), as shown in the
right panel of Fig.~\ref{fig:Diffuseone}.
A direct comparison of the nebular and continuum fluxes also dispels
the possibility that the observed trend is due to a sensitivity bias.
As discussed in Paper~V, our detection thresholds translate into a
rather narrow range for the minimum equivalent width of the emission
lines that we can measure, from 0.2\AA\ to just 0.07\AA\ for \Hb, so
that in practice also the flux of barely detectable \Hb\ lines scales
with the strength of the underlying stellar continuum.
Reassuringly, the right panel of Fig.~\ref{fig:Diffuseone} shows that
the integrated fluxes corresponding to such detection limits not only
lie comfortably below the total measured values of the \Hb\ flux but
also that they define a shallower correlation with the continuum
fluxes than found in the case of the observed \Hb\ fluxes.

The tight correlation between the integrated luminosity of the nebular
emission and that of the stellar continuum stems in reality from the
fact that within most of the galaxies in our sample the radial profile
of the ionised-gas emission follows very closely the stellar surface
brightness distribution.
Figure~\ref{fig:Diffusetwo} provides three specific examples (in the
case of NGC~2974, NGC~3414 and NGC~4150) that vividly illustrates how
well the nebular fluxes can follow the stellar continuum in early-type
galaxies, or alternatively, of how the equivalent width of \Hb\ can
take remarkably similar values at very different distances from the
galactic centre corresponding to very different continuum levels of
the stellar surface brightness (in contrast to the case of NGC~3032,
also shown in Fig.~\ref{fig:Diffusetwo}).
In fact, across our entire sample the equivalent width of the
\Hb\ line fluctuates on average by just 32\% around the mean value it
takes in any given galaxy, excluding objects with obvious star
formation (NGC~3032, NGC~4459, NGC~4526; \S\ref{subsec:Ostars}) and
while disregarding the central 3$''$ of all galaxies to avoid any
possible AGN contamination (\S\ref{subsec:AGN}).
This is conveyed also by Fig.~\ref{fig:Diffusethree} where the
\Hb\ and continuum fluxes contributed by each bin of our sample
galaxies (with the same exceptions) are plotted against each other,
after rescaling the \Hb\ and continuum fluxes observed in each galaxy
to common average values.
A power-law fit to such a stacked distribution indicates a nearly
linear relation ($a=0.88\pm0.01$) with 40\% scatter (0.14 dex).

\placefigDiffusethree

\subsubsection{Explaining the Nebular to Stellar Connection}
\label{subsubsec:Dehnen}

Does a nearly constant equivalent width for the recombination lines
necessarily imply a diffuse stellar photoionising source?
After all, the nebular flux and the stellar surface brightness, the two
elements entering the equivalent width definition, relate to two
different integrals of the stellar luminosity density.
Whereas the surface brightness is a line-of-sight integration of the
luminosity density, the recombination flux is connected to the total
Lyman continuum emitted by the putative ionising stellar
sub-population, which reaches the gas from all directions within the
galaxy.

The simplest model for the radial profile of the equivalent width of
\Hb\ can be computed in the case of a spherical galactic geometry and
assuming that a constant fraction of the ionising radiation from the
evolved stellar sources is absorbed and subsequently re-emitted across
the galaxy by the ionised-gas, where the latter is also taken to
reside in a planar configuration.
Under the previous assumptions the equivalent width of the
recombination lines at a given projected distance $R$ from the center
will be directly proportional to the ratio of total stellar radiation
$F(R)$ to the stellar surface brightness $\Sigma(R)$ at the same point,
since also the stellar ionising continuum reaching the gas clouds
should scale with the stellar radiation $F(R)$ (for a set stellar
population).
For any given intrinsic luminosity density profile $L(r)$ the integral
that delivers $F(R)$ is, in spherical coordinates,
\begin{equation}
   F(R)=\int_0^{\infty}{dr} \int_0^{2\pi}{d\phi} \int_0^{\pi}{\frac{L(r)}{4\pi x^2} r^2 \sin{\theta} d\theta}
\label{eq:starrad}
\end{equation}
where $x=\sqrt{R^2+r^2-2rR\cos{\theta}}$. Substituting
$\cos{\theta}=\mu$, $d\mu = -\sin{\theta}\, d\theta$ the previous
triple integral can be reduced to
\begin{eqnarray}
   F(R) & = & \int_0^{\infty}{\frac{L(r)r^2}{2}dr} \int_{-1}^{1}{\frac{d\mu}{R^2+r^2-2Rr\mu}} \\
        & =  & \int_0^{\infty}{\ln{\left(\frac{R+r}{R-r} \right)^2} \frac{r}{4R} L(r)dr}
\label{eq:starradbis}
\end{eqnarray}
Adopting a $\gamma$--model for the intrinsic luminosity density of
early-type galaxies \citep{Deh93}
\begin{equation}
   L(r)=\frac{(3-\gamma)M}{4\pi}\frac{a}{r^\gamma(r+a)^{4-\gamma}}
\end{equation}
where $M$ is the total luminosity (or stellar mass) of the galaxy and
$a$ is a scaling radius (which relates to $R_e$ through $\gamma$), the
integral given by Eq.~\ref{eq:starradbis} becomes
\begin{equation}
   F(R)=\frac{(3-\gamma)Ma}{16\pi R}
   \int_0^{\infty}{\frac{r^{1-\gamma}}{(r+a)^{4-\gamma}}\ln{\left(\frac{R+r}{R-r}\right)^2}dr}
\end{equation}
This can be numerically evaluated by breaking it in two pieces: from
$r = 0 \to R$ substituting $r = R u$, and from $r = R \to \infty$
substituting $r = R/u$. The $\gamma$--models also allow to solve for
the surface brightness profile $\Sigma(R)$ \citep[see Appendix B
  of][]{Deh93}, which can be varied to include fairly shallow models
($\gamma=1$, corresponding to Hernquist 1990 profile) 
as well as rather steep profiles ($\gamma=2$, corresponding to a Jaffe
1983 profile).

With both $F(R)$ and $\Sigma(R)$ at hand, Fig.~\ref{fig:Diffusefour} shows
that according to our simple model the equivalent width of the
recombination lines should increase towards the outskirts of
early-type galaxies, with little dependence on the choice of the
intrinsic $L(r)$ profile.
Between $\sim$10\% and $\sim$70\% of one effective radius $R_e$, which
on average correspond to $3''$ and the radius containing 90\% of the
emission in our sample galaxies, the model predicts a $\sim$80\% rise
in the equivalent width of \Hb.
Although a 0.25 dex increase may be accomodated by the average trend
observed across our sample (Fig.~\ref{fig:Diffusethree}), more often
than not the equivalent width of the \Hb\ line tend to remain constant
or even decrease with radius, in particular if we focus on the objects
with the most extended and diffuse \Hb\ emission (e.g.,
Fig.~\ref{fig:Diffusetwo}) and exclude those where the impact of other
sources of ionisation than evolved stars may be particularly important
(e.g., the filaments of NGC~4486).

\placefigDiffusefour
\placefigDiffusefive

Can such a discrepancy be remedied by relaxing the simple premises of
our model? An important aspect to keep in mind is that not all the
ionising photons emitted by the evolved stellar sources may reach the
gas due to interstellar absorption, so that only the stellar radiation
emitted within the mean-free path $l$ of the photons should be
considered. This will lead to a steeper $F(R)$ profile and a shallower
$F(R)/\Sigma(R)$ gradient, in better agreement with our data. Yet, as $l
\to 0$ and $F(R) \to L(r) \cdot l$ the $F/I$ ratio will eventually
decrease as fast $R^{-1}$, again in disagreement with our
observations.
As regards our assumption that a constant fraction of the ionising
photons is absorbed across the galaxy, this depends on how closely the
radial profiles of the gas column density and of ionising radiation
follow each other, and on whether the filling factor of the gas clouds
varies.
The column density of a steady disk increases towards the center of a
realistic galactic potential well, as the models of \cite{Kaw07}
illustrate. The work of Kawata et al. (2007) also shows that outside
the very central regions the Toomre parameter $Q$ is rather constant
(see their Fig.~2), so that the gas surface density profile
$I_{gas}(R)$ should scale almost exactly with the epicyclic frequency
$\kappa$ (given that $Q \propto \kappa/I_{gas}$). Using the
$\gamma$--models to compute $\kappa$ one learns that the gas surface
density decreases always more gently than $F(R)$ with radius, and even
more so when compared to $L(r)$. For a fixed filling factor it is
therefore likely that an increasing number of ionising photons escape
the galaxy as we approach the centre, whereas towards the outskirts of
the gaseous disk an increasing fraction of the gas may remain neutral.
The combination of a small mean-free path $l$ and of a flattening of
the \Hb\ flux profile towards the centre (as it departs from the $F(R)
\sim L(r)$ profile) could thus provide a relatively simple explanation
for the general behaviour observed across our sample galaxies of a
nearly constant equivalent width of \Hb.
Alternatively, if the ionising photons travel far enough in the galaxy
that $F(R)$ is not significantly affected by $l$, another plausible
avenue to explain our data is to consider the possibility that the
ionising sources are more concentrated than the bulk of the stellar
populations.
This may be the case if the ionising stars are also responsible for
the far-UV flux ($\sim$ 1500\AA) observed in early-type galaxies. The
far-UV light displays indeed a steeper surface-brightness profile than
observed at optical wavelengths, following closely also the
metallicity gradient traced by the Mg$b$ line-strength index
\citep{Jeo09}.

\subsubsection{Ionisation Balance}

In the absence of a definite model for the tight connection between
the values of the stellar and nebular luminosity of early-type
galaxies, we turn our attention to the energy budget. Specifically, we
ask whether the various ionising stellar candidates can provide a
sufficient number of ionising photons to explain the observed
\Hb\ fluxes.
In doing so we return to the use of integrated flux measurements,
since within large apertures the stellar luminosity estimated from the
integrated flux approximates well the total luminosity within the
galaxy, and therefore provides a basis for estimating the total
ionising stellar radiation.

As mentioned above, \citet{Mac96} already carried out the
ionisation-budget exercise in the case of pAGB stars. 
Using the predictions of \citet{Bin94} for the specific ionising
photon luminosity from such stars (that is, per unit mass of the
entire stellar population) and assuming that all the ionising
radiation is re-processed into nebular emission, \citeauthor{Mac96}
could convert the stellar luminosity within the average radius of the
emission-line regions into a limiting value $L_{\rm H\alpha, pAGB}$
for the \Ha\ luminosity that could be possibly powered by pAGB stars.
Here we revisit \citeauthor{Mac96} experiment, using the stellar
population models of \citet{Yi99} to compute the Lyman continuum from
pAGB stars and 2MASS K-band images for our sample galaxies to trace
the stellar mass within the radius containing 90\% of the nebular
emission. Consistent with previous work, the specific number of pAGB
stars in the models of \citeauthor{Yi99} is pretty much constant
between 4 and 11 Gyr, which brackets the typical age of the stellar
population dominating the stellar mass of our objects.
Assuming case B recombination the specific Lyman luminosity of pAGB
stars in these models would deliver 0.007 ${\rm erg\, s^{-1}}$ in
\Ha\ photons for each K-band ${\rm erg\, s^{-1}}$, if all ionising
photons are intercepted and re-processed. Under these circumntances,
Figure~\ref{fig:Diffusefive} shows that pAGB stars still provide an
abundant number of ionising photons to explain the ionised-gas
emission observed in our sample galaxies. That the predicted limiting
values $L_{\rm H\alpha, pAGB}$ far exceed the observed values $L_{\rm
  H\alpha}$ is encouraging considering that for many objects the
emission-line regions cover only a fraction of the area used to
compute such upper-limits (shown with a whiter shade of grey), so that
almost certainly a good fraction of the pAGB ionising radiation
escapes the galaxy.
On the other hand, we need to keep in mind that the present stellar
population models may be grossly overestimating the specific number of
pAGB stars, and therefore their ability to power the ionised-gas
emission in early type galaxies. The recent near- and far-UV
observations of M32 with the Hubble Space Telescope by \citet{Bro08}
suggests indeed a dearth of pAGB stars compared to standard
predictions for their numbers.

Turning our attention to other stellar candidates, we now consider the
role low-mass X-ray binaries (LMXBs).
The exquisite spatial resolution of the Chandra X-ray observatory has
recently allowed to detect LMXBs in nearby ellipticals, enabling the
derivation of their luminosity function and, most important in the
present context, their total X-ray luminosity \citep[in the 0.3-8 keV
  range; e.g.][]{Kim04}. According to \citeauthor{Kim04} the LMXB
population within the optical radius of their host galaxies can
deliver from a few times $10^{39}$ to several $10^{40}{\rm erg\,
  s^{-1}}$.
The X-ray spectral energy distribution of LMXBs in early-type galaxies
has been found to be well characterised by a simple power-law with
photon index $\Gamma = 1.56$ \citep{Irw03}, corresponding to an energy
index $\alpha = 0.56$ (for $F_\nu \propto \nu^{-\alpha}$).
For a power-law spectrum with an energy index ${\rm \alpha=0.1}$ to
$0.9$ the \Ha\ luminosity produced by reprocessing the X-ray emission
can range from $1/9$ to $1/2$ of the X-ray luminosity \citep{Ho08}.
Thus, to a first approximation, the overall LMXBs population of
early-type galaxies could provide enough ionising photons to power the
observed nebular emission, given that the \Ha\ luminosity of the
latter ranges between a few times $10^{38}$ and several $10^{39}{\rm 
erg\, s^{-1}}$ (see Paper~V).
Encouraged by this first check we used the relation found by
\citeauthor{Kim04} between the K-band luminosity $L_{K}$ of early-type
galaxies and the X-ray luminosity $L_{\rm X,LMXBs}$ of their LMXB
population to more accurately estimate for each of our sample galaxies
the limiting \Ha\ luminosity $L_{\rm H\alpha, LMXBs}$ that could be
powered by these stellar remnants.
Upon such a closer inspection, and assuming that 18\% of the LMXBs
X-ray luminosity converts into \Ha\ emission,
Fig.~\ref{fig:Diffusefive} shows that LMXBs do not live up to our
expectations as ionising sources in early-type galaxies. In
particular, Fig.~\ref{fig:Diffusefive} reveals that although the
$L_{\rm H\alpha, LMXBs}$ values are on average consistent with the
observed \Ha\ luminosities, in agreement with our previous estimates,
in practice none of our sample galaxies could display emission powered
solely by LMXBs if the gas coverage is taken into account.
In fact, in all objects with $L_{\rm H\alpha, LMXBs} \ge L_{\rm
  H\alpha}$ in Fig.~\ref{fig:Diffusefive} most the LMXBs ionising
radiation is likely to escape the galaxy because of a patchy gas
distribution, whereas when the gas coverage is high the LMXB
population seems unable to power the nebular emission.

Finally we explore the case of extreme horizontal-branch stars (EHBs).
Such low-mass core Helium burning stars only appear in significant
numbers in old stellar systems, contributing in particular to the
far-UV flux in populations with ages greater than 9 Gyr. For this
reason, EHBs have been suggested as the dominant source for the
UV-upturn observed in many early-type galaxies \citep[\eg][]{Gre90}
and as indicators of the oldest stellar populations \citep{Yi99}.
Although EHBs are much weaker as ionising sources than pAGB stars,
their contribution to the Lyman continuum becomes nonetheless
significant in stellar populations older than 10 Gyr, eventually
overtaking pAGBs as the main source of ionising photons when the
population turns 12 Gyr old.
In order to trace the contribution of EHBs in our sample galaxies we
used GALEX measurements for the far-UV flux (1341--1809\AA), extracted
within the area containing 90\% of the nebular emission as in the case
of our 2MASS measurements.
Contrary to the case of the near-UV passband (centred at
$\sim$2310\AA), the far-UV measurements are fairly insensitive to the
contribution of younger stellar components (i.e. from few 100 Myr to 1
Gyr), whereas pAGB stars do not contribute to either of these
passbands.
To transform the GALEX far-UV fluxes into a limiting \Ha\ luminosity
from EHBs, $L_{\rm H\alpha, EHBs}$, we used the models of
\citeauthor{Yi99} to integrate the Lyman continuum due only to stars
either on horizontal-branch or that have only recently evolved from
it, which includes also AGB-manqu\'e stars.
We adopted models for a 11-Gyr-old stellar population, which reproduce
the UV-to-optical colour observed in prototypical UV-upturn galaxies
such as NGC~4552. Metallicity does not impact on our results as this
parameter only produces differences in the models in UV regions ($\sim
1100$\AA) outside the far-UV passband.
In this framework the models of \citeauthor{Yi99} yield 0.0295 ${\rm
  erg\, s^{-1}}$ in \Ha\ photons for each far-UV ${\rm erg\, s^{-1}}$.
As in the case of LMXBs, Fig.~\ref{fig:Diffusefive} shows that EHBs
per se are unlikely to generally power the nebular emission observed
in early-type galaxies.

\subsubsection{Summary on the Role of Evolved Stellar Sources}

To conclude, we have added to early suggestions that the main source
of ionisation for the gas of early-type galaxies is to be found among
the bulk of their stellar population.
We have found a very tight correlation between the total values of the
nebular and stellar luminosity in our sample galaxies, when
considering only the regions where the ionised-gas emission is
detected. Furthermore, we have shown that such a global trend stems
actually from the fact that the ionised-gas emission of early-type
galaxies follows very closely their stellar surface brightness.
Although we have not been able to conclusively explain how such a
constant profile for the equivalent width of the nebular emission can
be produced by stellar photoionisation, through simple modelling and
reasoning we have shown the direction for future investigations
without invoking embarassing fine tuning of key parameters such as the
mean free path of the ionising photons.
Overall, pAGBs remain the favourite ionising candidates to explain the
gas emission of early-type galaxies, which points to a need for
further investigations into the present discrepancy between
observations and models as regards their total numbers.
The role of LMXBs would seem marginal in this context, unless future
observations reveal a population of super-soft X-ray sources, which
would ionise more efficiently the gas surrounding them and induce a
very specific nebular spectrum \citep{Rap94}.
In principle, the relative importance of pAGBs, LMXBs and EHBs could
be further investigated through a careful emission-line diagnostic,
since this analysis can trace the specific shape of the ionising
continuum of the different sources. For instance, EHBs have a much
softer Lyman continuum compared to pAGBs, which should make it more
difficult for EHBs to produce extended partially-ionised regions.
Unfortunately a diagnostic analysis based on the \Ni\ emission is not
possible for the diffuse component discussed here, since the
\Ni\ lines are detected almost exclusively in the central regions of
our galaxies, in the presence of star formation, or in the filamentary
structures of the most massive of our sample galaxies where the
interaction with the hot ISM is the most likely source of ionisation
(\S\ref{subsec:Xray}).
Notwithstanding the absence of diffuse \Ni\ emission, the behaviour of
the \Oiii/\Hb\ ratio is still useful in this context. Specifically, we
note that strong gradients and peculiar features are {\it not\/}
expected to arise easily in the presence of a uniform radiation field
from the same kind of source across the galaxy, unless to consider
contrived structures of higher dust concentration or enhanced gas
number density that could allow for local fluctuation of the ionising
parameter $q$. The finding of remarkable \Oiii/\Hb\ structures
suggests therefore the role of additional ionising mechanisms, such as
those discussed in the next sections.

\subsection{Shocks}
\label{subsec:Shocks}

\subsubsection{Fast Shocks}

Shocks offer a natural explanation for some of the most prominent and
coherent structures that are observed across our sample in the maps
for the equivalent width of the \Hb\ and \Oiii\ lines and for the
\OiiioHb\ ratio, in particular for morphologies like spiral arms, oval
and integral-sign shaped structures (\eg\ Fig.~\ref{fig:Shockone}).
Such features can indeed be induced by the presence of stellar bars or
weak triaxial perturbation of the stellar potential, which can funnel
the gas into preferential streams where 
fast (above 100 \kms)
shocks between gas clouds may occur.
Yet, the presence of equivalent-width structures per se does not
guarantee that shocks are powering the gas emission - other diffuse
sources (such as pAGB stars) may be still ionising the gas as this
accumulates in spiral arms and similar features, where the observed
emission would be stronger only due to a higher column density.
On the other hand, the case for shock excitation becomes much stronger
when the features in the \Oiii\ or \Hb\ emission are accompanied by
similar structures in the \OiiioHb\ ratio.
Similarly, we should expect a correlation between the strength of the
recombination \Hb\ emission and the velocity dispersion of the lines
in the presence of fast shocks \citep[see, e.g.,][]{Dop95}, because
the ionising flux powered by the latter scales with the shock velocity
and because unresolved velocity gradients could in principle be traced
by an enhanced velocity dispersion.

\placefigShockone

Associated \OiiioHb\ structures are more often observed for
integral-sign emission features (where the \OiiioHb\ distribution can
be oddly asymmetric, \eg\ NGC~4262, NGC~4278 and NGC~4546), and only
in NGC~3414 in the case of spiral arms. As regards the gas velocity
dispersion \Sg, this quantity does not seem to generally trace notable
equivalent-width features, except for NGC~4546 
where \Sg\ appears to display larger values along the locus of the
main integral-sign structure.
That remarkable equivalent-width structures are not always accompanied
by similarly obvious \OiiioHb\ and \Sg\ features 
%
(as Fig.~\ref{fig:Shockone} illustrates for \Sg)
%
suggests that shocks may indeed not always be the main source of
ionisation in these regions.
On the other hand, the shock geometry may not lead to significant
broadening (as set by the spectral resolution of the \sauron\ data) of
the emission lines along the line of sight, for instance if the shocks
occur in regions that are much smaller than the physical scales that
we can resolve (typically $\sim$100pc) and if they are surrounded by a
more relaxed ionised-gas medium.
In this respect, it is interesting that all galaxies featuring notable
spiral arms or integral-sign structures display also extended central
velocity dispersion peaks 
%
(see also Fig.~\ref{fig:Shockone}).
%
Such central \Sg\ gradients could mark the point where the previous
limitations are overcome and the kinematic signature of the shocks can
be read, if the shock fronts wind up and get closer together as the
gas is funneled towards the center.
This certainly appears to be the case for NGC~2974, where
\citet{Ems03} showed with high-resolution integral-field spectroscopic
observations that the spiral-arm morphology observed in the
\sauron\ gas maps extends well towards the centre. Emsellem et
al. could also resolve better the central \Sg\ gradient in the
\sauron\ data, finding \Sg\ peaks on the inner sides of the central
spiral arms that further suggest the presence of shocks, although yet
not necessarily that these can power the ionised-gas emission.

When \Ni\ emission is detected in objects with notable spiral or
integral-sign features, the corresponding \Ni-diagnostic analysis is
always consistent with shock-excitation. The \Ni-diagnostic is seldom
conclusive, however, since most of the \Ni\ emission originates close
to the centre of our sample galaxies (see e.g., the cases of NGC~2974
and NGC~4278 in Fig.~\ref{fig:Shockone}) where the \Ni-diagnostic
diagrams indicate that also AGN activity could be powering the gas
emission.
In fact, if shocks are responsible for the gas excitation in these
regions, the grids of Fig.~\ref{fig:Shockone} suggest exceedingly high
shock velocities between 300 and 500 \kms, which can hardly be
explained through gravitational motions.
For instance, the gas-dynamical models of \citet{Ems03} for the inner
regions of NGC~2974 suggest gas streamlines crossing at an angle of
just 45$^{\circ}$ (see their Fig.~21), which even assuming gas motions
at nearly circular speed (up to $\sim$300 \kms\ within the central
2-3$''$ according to the models of Paper~IV) would not allow to reach
the required shock velocities.
It thus seems unlikely that shocks are powering the nebular emission
observed in the central \Ni-emitting regions of these two galaxies.

Despite the previous shortcomings shocks could still have a role in
powering the gas emission observed in integral-sign features when such
structures are accompanied by notable asymmetric
\OiiioHb\ distributions, as is the case for NGC~4262, NGC~4278 and
NGC~4546 (see Fig.~4 of Paper~V).
Excluding AGN photoionisation for such extended structures and
star-formation, which usually proceed in relaxed morphologies such as
rings or disks, explaining such disparate \OiiioHb\ values through
photo-ionisation by a diffuse source such as pAGB stars would require
different values for the ionising parameter $q$ at the opposite sides
of a galaxy and therefore having either different values for the gas
number density or for the concentration of dust.
Maintaining such different physical conditions over time would seem
unlikely at first, which leaves us to consider the role of shocks.
One way to explain an asymmetric distribution for the \OiiioHb\ ratio
is to suppose that shock ionisation proceeds under different
conditions ahead and behind of the shock fronts and that dust
precludes us from receiving most of the nebular emission from either
side of such fronts, which will naturally occur when looking at one
end or the other of the galaxy. In fact, shock models such as those of
\citet{Dop95,Dop96} feature the possibility of adding a gaseous
precursor ahead of the shocks, which would naturally occur in gas-rich
galaxies. Such precursor \Hii\ regions are characterised by high
ionisation, which makes them efficient \Oiii\ emitters and therefore
good candidates as the principal source of nebular emission along the
line of sight where the highest \OiiioHb\ values are observed.
Figure~\ref{fig:Shockone} shows the predictions also of the shock
models with a gaseous precursor, which indeed can explain the observed
values for the \OiiioHb\ ratio that typically range between $\sim$ 2
and 4. Yet, also in the presence of a precursor the model grids imply
rather large high $V_s$ values between 200 and 300\kms, which
admittedly could still be difficult to achieve, in particular in
objects with well organised gas motions like NGC~4278.

In fact, our premise to discard the role of a diffuse source of
ionisation in order to explain such asymmetric \OiiioHb\ structures,
that an uneven dust distribution would be unlikely, is actually
unfounded according to Spitzer observations.
Based on IRAC images, in \citeauthor{Sha09} we uncover features in
maps for the 8$\mu$m non-stellar emission excess that neatly
correspond to the regions in NGC~4278 where the \OiiioHb\ ratio falls
at its lowest value. Such a 8$\mu$m flux excess is usually associated
across our sample to star-forming regions, and comes from ionised-PAH
emission. In NGC~4278 there is no evidence for on-going star
formation, however, in particular given the absence of a substantial
molecular reservoir\footnote{The single-dish detection of
  \citet{Com07} in NGC~4278 is only marginal compared to other
  galaxies showing evidence for on-going (e.g., NGC~3032, NGC~4459 and
  NGC~4526) or recent (e.g., NGC~3156 and NGC3489, see \S
  \ref{subsec:HIERs}) star formation.}.
The 8$\mu$m excess observed here may be due to predominantly neutral
PAHs that emit more significantly in the 11.3 and 12.7$\mu$m bands,
similarly to other early-type galaxies with PAH emission and no CO
detection \citep[see, e.g., the case of NGC~2974 in][]{Kan05}.
Such an asymmetric concentration of dust would induce a local
hardening of the Lyman continuum from evolved stellar sources in the
galaxy and would reduce the rate at which the ionising photons hit the
gas clouds, thus leading to a smaller ionising parameter $q$
\citep[for a discussion on dust effects see][and references
  therein]{Shi95}. Both effects would produce lower values for the
\OiiioHb\ ratio in regions with the highest dust covering.

Finally, shocks would also spring to mind as the most likely source of
ionisation for the emission observed in galaxies like NGC~3156 and
NGC~3489, when glancing at the extreme range of values for the
\OiiioHb\ ratio (from $\sim$1 up to 10) that is found across the
entire \sauron\ field of view in these objects (see
Fig.~\ref{fig:HIERsone}).
Yet, the high \OiiioHb\ values observed in these objects are
inconsistent with shock excitation if we consider that NGC~3156 and
NGC~3489 are among the least massive galaxies in our sample. The
circular velocities in these objects peak at just $\sim$150 \kms,
which would allow only for a restricted range of shock velocities (say
$V_s \le 200$ \kms) and thus only for a rather limited range of values
for the \OiiioHb\ ratio (up to 2, see Fig.~\ref{fig:Shockone}),
irrespective of the presence of a gaseous precursor.
Furhermore, NGC~3156 and NGC~3489 display fairly regular gas
kinematics that indicates the presence of a relaxed gaseous system,
which rules out the possibility of strong non-circular motions to
circumnvent the previous impasse with shock excitation.

\subsubsection{Slow Shocks}

The dynamical deadlocks encountered in the previous subsection may
suggest that slow shocks \citep[e.g.,][]{Shu79}, instead of fast
shocks, could be important. Slow shocks could be driven by ram
pressure into gas clouds as they move through the hot halo gas, and
have the appealing property of producing relatively strong
low-ionisation lines such as \Oi\ or \Ni, and values of the
\NioHb\ ratio close to what we observe.
Even assuming that all early-type galaxies can retain a halo of hot,
X-ray emitting gas, which is far from clear to be the case
\citep[e.g.,][but see also \S\ref{subsec:Xray}]{Mat03}, a scenario
involving slow shocks would run into two sorts of complications.

First, the values of the \OiiioHb\ ratio predicted by the
\citeauthor{Shu79} models are quite sensitive upon the shock velocity,
so that in practice slow shocks would be consistent with our data only
for a rather restricted range of shock velocities, between 85 and 95
\kms.  Such a condition seems unlikely to be generally satisfied,
however, considering the wide range of velocities at which the gas
clouds may be orbiting within our sample galaxies.

Second, if slow shocks were generally occurring in early-type
galaxies, they could lead to a profile for the equivalent width of the
recombination lines that would be inconsistent with the flat profiles
that we observe, and even more than what we simply estimated in the
case of photoionisation by old stars (\S\ref{subsubsec:Dehnen}).
If the local density of the hot gas scales like the square root of the
stellar density \citep[as in giant ellipticals,][]{Tri86} we can
compute the radial trend that the emission from slow shock would have,
using the same simple assumptions and models of
\S\ref{subsubsec:Dehnen} and considering that the line emission from
such shocks scales like the cube of the orbital velocity of the clouds
(presumably close to circular) and the density of the hot halo gas.
Within the same region between 10\% to 90\% of an effective radius
that we considered in \S\ref{subsubsec:Dehnen}, the equivalent width
of emission arising from slow shocks would increase markedly a factor
2.5, which is not supported by observations.
Furthermore, the equivalent width of lines arising from slow shocks
would drop to zero towards the centre, posing another inconsistency
with the data in the absence of additional central source such as an
AGN.

\placefigHIERsone

\subsubsection{Summary on the Role of Shocks}

%
To conclude, shocks, either fast or slow,
can hardly be regarded as an important source of ionisation for the
diffuse nebular emission observed in early-type galaxies.
%
In particular,
we have discussed how fast shocks could contribute to power the
ionised-gas emission in regions where the gas is funneled into spiral
and integral-sign structures by the non-axisymmetric perturbations of
the gravitational potential. Yet, even there the role of diffuse
stellar sources may be as, if not more, important than shocks, in
particular when the spiral and integral-sign features in maps for the
\Oiii\ or \Hb\ emission are not accompanied by similar
\OiiioHb\ structures.
In the central regions of our sample galaxies shocks could contribute
to explain the high values for the gas velocity dispersion, but
limited \Ni\ diagnostic analysis combined with simple dynamical
arguments appear to dismiss shocks as the main source of gas
excitation.
Arguments for the importance of shocks, which included also the
presence of a gaseous precursor and of dust, were put forward also to
explain the peculiar asymmetric \OiiioHb\ distribution that is often
associated to integral-sign emission structures, but ultimately run
into a similar dynamical dead lock. In these cases Spitzer
observations reveal an uneven distribution of dust, which can help
reconcile the observed \OiiioHb\ character of the nebular emission
with simple photoionisation by diffuse stellar sources.
Finally, we have also considered the case for shock excitation in
low-mass galaxies with high-ionisation extended regions, and
eventually ruled out shocks as the main source of ionisation. In the
next section we will consider an alternative explanation for such
extreme cases.

\subsection{Post Starbursts and High-Ionisation Extended Regions}
\label{subsec:HIERs}

An intriguing subset of the early-type galaxies surveyed in the course
of the \sauron\ survey consists of relatively small objects (with
$\sigma_e \le 120$\kms) displaying high-ionisation extended regions
(HIERs), where the \OiiioHb\ ratio reaches some of the most extreme
values observed in our sample with great variations across the
\sauron\ field of view (Fig.~\ref{fig:HIERsone}). NGC~3156 and
NGC~3489, which we already considered in \S\ref{subsec:Ostars} and
\S\ref{subsec:Shocks}, represent only the most remarkable examples of
such a kind of early-type galaxies - HIERs are also found in NGC~7332
and NGC~7457 (see Fig~4 of Paper~V).
A common trait that these low-mass objects share is the presence
throughout the galaxy of fairly strong \Hb\ absorption features
(Paper~VI), which is indicative of a recent star-formation episode.
In fact, star formation in NGC~3156 and NGC~3489 may have not yet come
to a complete halt, in particular in those relatively small regions
where very low values of the \OiiioHb\ ratio are observed. Both PAH
and CO reservoirs have been detected in these galaxies \citep[see
  Shapiro et al. 2009 and ][respectively]{Com07}, although the amount
of molecular material found in NGC~3156 and NGC~3489 ($M_{\rm H2} =
0.2, 0.1 \times 10^8 \rm M_{\odot}$) is significantly less than what
found in the presently star-forming galaxies NGC~3032, NGC~4459 and
NGC~4526 (with $M_{\rm H2} = 2.5, 1.7, 3.7\times 10^8 \rm M_{\odot}$).
Additionally, the gas kinematics in NGC~3156 and NGC~3489 trace a
remarkably cold dynamical system (the values of the gas velocity
dispersion \Sg\ observed in NGC~3156 and NGC~3489 among the smallest
measured in our sample, $\sim$50 \kms) which could still favour the
formation of stars.
All these facts suggest that these objects are post-starbursting
systems, perhaps linked to ``E+A" galaxies\footnote{The very nuclear
  regions, within the few tens of pc, of NGC~3489 already display the
  classical features of an E+A spectrum \citep{Sar05}.} \citep{Dre83}.

Is the high-ionisation character of the nebular emission from these
objects linked to their recent star-formation history?
In the context of AGN activity, \citet{Tan00} considered the role of
post-starbursting stellar systems to explain the LINER and Seyfert
emission of nearby galactic nuclei. Namely, Taniguchi et
al. recognised that as the most massive and recently formed stars
start leaving the asymptotic giant branch ($\sim 10^8\rm yrs$ after
the starburst), the corresponding planetary nebulae nuclei become an
important and sufficiently hot source of ionising photons.
Whereas this scenario is fairly limited in explaining AGNs , in
particular since most galactic nuclei are invariably old
\citep[\eg][but see also \citealt{Ho08}]{Sar05}, photo-ionisation from
the pAGB population associated to a recent starburst is a very likely
ionising mechanism for the HIER emission observed in low-mass
early-type galaxies.

According to Taniguchi et al., to power with pAGB stars from massive
progenitors (between 3 to 6 $\rm M_\odot$) the \Ha\ luminosity of
NGC~3156 and NGC~3489, which is around $\sim 10^{39} \rm erg\,
s^{-1}$, these objects should have experienced a starburst involving
approximately $\sim 10^9 \rm M_\odot$ of gas, this is consistent with
our estimates for the amount of stars that should have recently formed
in these galaxies (corresponding to 10\% of the total mass) in order
for the values of their stellar and dynamical mass-to-light ratio to
agree with each other as is the case for more quiescent early-type
galaxies in the \sauron\ sample (Paper~IV).

Another attractive feature of such a post-starburst photo-ionisation
scenario is that it could naturally account for the extreme values of
the \OiiioHb\ ratio observed in early-type galaxies with HIERs.
The pAGB population associated with the recent starburst is presumably
still confined to the galactic plane where star formation took place
and where the remaining gas is still orbiting, thus bringing the gas
much closer to its photoionising sources than is the case for the pAGB
stars related to the older bulge population (\S\ref{subsec:OldStars}).
In turn this situation would translate into an higher ionisation
parameter $q$ and therefore to larger values for the \OiiioHb\ ratio.
We note also that due to their proximity to the gas, the Lyman
continuum of the pAGB stars associated to a recent starburst could be
reprocessed as nebular emission more efficiently than for other
stellar sources throughout the bulge. This would help explaining also
the remarkable strength of the nebular emission in objects like
NGC~3156 and NGC~3489, where the equivalent width for the integrated
\Hb\ and \Oiii\ lines reaches values of 0.5\AA\ and 1.5\AA,
respectively.\\


To conclude, in approximately 10\% of the early-type galaxies in the
\sauron\ sample the nebular emission is most likely powered by pAGB
stars associated to a recently formed population of stars. The intense
and highly-ionised character of the observed emission in these objects
can be explained by the vicinity of the gas to its ionising stellar
sources, which are presumably still confined to the galactic plane
where they formed.
The nature of the nebular activity observed in the post-starbursting
systems of the \sauron\ sample could also have bearings for our
understanding of the ``E+A'' phenomenon. ``E+A" galaxies are extremely
rare objects \citep[e.g.,][]{Kav07} that display strong Balmer
absorption lines but no nebular emission, suggestive of a recent but
already exhausted, perhaps even quenched, star-formation episode.
Yet, ``E+A" are not totally devoid of gas since they often still
display abundant \Hi\ reservoirs \citep{Buy06}, which, according to
our picture, could be re-ionised just a few $\sim 10^8\rm yrs$ after
the starburst. This would leave only a very short time after the death
of OB-stars for the galaxy to appear quiescent while developing A-star
type features, which could contribute to explain why so few
post-starbursting galaxies are observed in their ``E+A" phase.

\placefigXrayone
\placefigXraytwo

\subsection{Interaction with the Hot Interstellar Medium}
\label{subsec:Xray}

Historically, it was perhaps the finding of optical emission in and
around the most massive galaxies of X-ray selected clusters that first
drew considerable attention to the ionised-gas component of early-type
galaxies \citep{Hu85, Hec89}.
The discovery of such a warm medium was indeed regarded as a major
success of the cooling flow theory \citep{Fab94}, although it soon
became apparent that the amount of ionised gas found in the optical
filaments of X-ray clusters exceeded by far the rates predicted by
cooling flows \citep{Hec89}.
Furthermore, the presence of dust in such filaments posed an
additional problem to the cooling flow scenario since gas cooling down
from a KeV plasma should not be dusty and since dust particles ought
to be quickly destroyed by X-ray sputtering in environments permeated
by hot ions \citep[e.g.,][]{Don93}.
The connection between the X-ray and optical emission in and around
the most massive early-type galaxies remained an established fact,
however, which led to the suggestion that the nebular emission arises
in fact from material that has been recently accreted and that is
being excited by the hot medium itself, either through thermal
electron conduction \citep{Spa89,deJ90} or by the radiation that the
hot gas emits as it cools down \citep{Voi90}.
Sparks et al. (1989) showed that the energy flux provided from heat
conduction is sufficient to explain the nebular emission, whereas
\citet{Don91} found that the X-ray gas radiation could explain also
the typical line ratios observed in the optical filaments of cooling
flows clusters \citep[see also][for a recent investigation]{Fer09}.
Yet, the most spectacular confirmation of the connection between the
hot and warm phases of the ISM came when X-ray images of high spatial
resolution revealed a striking spatial coincidence between regions
displaying optical and X-ray emission, first in the case of M87
\citep{You02,Spa04} and then in NGC~5846 \citep{Tri02}.

In light of these findings we turn our attention to the interaction
between the warm and hot phases of the ISM, as the last ionising
mechanism that we will consider in order to explain the nebular
emission observed across our own sample of early-type galaxies.
Unsurprisingly, it is only the most luminous galaxies in the
\sauron\ sample that can retain a massive halo of hot, X-ray emitting
gas, although the degree of rotational support also appears to play a
part in determining the presence of an X-ray atmosphere.
Figure~\ref{fig:Xrayone} vividly illustrates this point by showing how
only the brightest galaxies feature values of the X-ray luminosity
\citep[mostly from the compilation of][]{oSu01} that well exceed what
is expected from an unresolved population of LMXBs, in particular if
considering slowly- or non-rotating galaxies (green or blue circles).
All but two (NGC~4552, NGC~5982) of such massive systems display
prominent dust features (see the HST maps of Paper~V), and luckily
Chandra images are available for these dusty giant galaxies (NGC~4374,
NGC~4486, NGC~5813, NGC~5846) in order for us to investigate the
connection between the X-ray and optical emission.

Figure~\ref{fig:Xraytwo} compares the Chandra X-ray images for the
four brightest slowly- or non-rotating galaxies in the \sauron\ sample
with maps for their \Hb\ emission from Paper~V. The Chandra images
show integrated fluxes in the 0.3 -- 5.0 keV energy band and 
have been spatially rebinned through Voronoi tesselations 
\citep{Cap03} to reach a minimum of 10 counts per bin.
Except in only a few regions, the ionised-gas emission is usually 
associated to X-ray emitting features, whereas the converse does not 
always hold.
Thus, with NGC~4374 and NGC~5813 we bring two additional examples
besides NGC~4486 and NGC~5846 of objects displaying such a remarkable
coincidence between the hot and warm phases of the ISM.
\citet{Spa04} could also show that the X-ray filaments of NGC~4486
have a lower temperature than the surrounding medium, supporting the
idea that heat is being transferred from the hot to the warm gas.
Unfortunately the Chandra images for the other objects in
Fig.~\ref{fig:Xraytwo} are not sufficiently deep to allow the same
kind of analysis, since binning up to reach the minimum signal quality
for a sensible spectral analysis (usually 50 counts) would leave only
a few patches.
Yet, our integral-field data reveal another characteristic of the
ionised-gas emission in such dusty giant slowly- or non-rotating
galaxies that sets them aside from the rest of the \sauron\ sample and
which reinforces the case for gas excitation through the interaction
with the hot medium.

\placefigXraythree

Figure~\ref{fig:Xraythree} shows indeed how the optical filaments of
such dusty giant ellipticals display rather intermediate values for
the \OiiioHb\ ratio ($\sim$1--1.5), which lie below the average
\OiiioHb\ values observed in the other quiescent early-type galaxies
in the \sauron\ sample.
This low-ionisation pattern is reminiscent of the low \OiiioHb\ values
that are observed in the optical filaments in and around the brightest
cluster galaxies \citep[e.g.][]{Sab00,Hat06}. Yet, in such
environments the \OiiioHb\ ratio falls to much lower values than we
observe in our sample galaxies when optical and X-ray emission
coincide.
For instance, \citet{Hat06} hardly detect any \Oiii\ emission in the
filaments that extend well beyond the optical boundaries of NGC1275,
the brightest member of the Perseus cluster, whereas in the inner
region of the same ionised-gas nebula \citet{Sab00} measure
\OiiioHb\ $\sim$ 0.3--1, still below the values that we find in the
four galaxies of Fig~\ref{fig:Xraytwo}.
Rather than considering varying Oxygen abundances, \citeauthor{Hat06} 
interpret this behaviour as an ionisation gradient driven by a progressive 
intensification of the excitation mechanism toward the optical regions of 
NGC1275.
We agree with this conclusion, and further suggest that the
intermediate \OiiioHb\ values observed in the optical regions of our
sample galaxies where hot and warm gas appear to interact are in fact
caused by the contribution of stellar photoionisation, in addition to
the interaction with the hot ISM.
There is indeed no reason to exclude in dusty giant ellipticals the
presence of a Lyman continuum from old stars, which per se would lead
to the larger \OiiioHb\ values that are observed in the rest of the
quiescent early-type galaxies of the \sauron\ sample.
In fact, we note that for the optical filaments associated to X-ray
features the radial profile of the \Hb\ flux tends to follow rather
closely the stellar surface brightness, further suggesting the
importance of stellar photoionisation.\\


To conclude, in the brightest and slowly- or non-rotating early-type
galaxies the presence of a massive halo of hot, X-ray emitting gas
imprints a distinct low-ionisation character to the emission-line
regions beyond the direct influence of a central AGN.
This line-ratio pattern is achieved thanks to the energy input
provided by the hot gas as it interacts, presumably through heat
conduction, with the warm gas, which is still subject to
photoionisation from old stellar sources like in other early-type
galaxies.
As reported in previous work, also in the dusty giant galaxies of the
\sauron\ sample a remarkable spatial coincidence between optical and
X-ray emitting features testify to the interaction between the hot and
warm phases of the ISM.

\section{Implications for Large-Scale Surveys}
\label{sec:SDSS}

If the high sensitivity of the \sauron\ survey has revealed the extent
to which diffuse ionised-gas emission is found in nearby early-type
galaxies (up to $\sim75\%$, Paper~V), shallower but much larger
spectroscopic campaigns such as the Sloan Digital Sky Survey (SDSS)
can help reveal through shear numbers important connections between
the nebular and stellar properties of early-type galaxies
\citep[e.g.,][]{Kau03,Gra07,Sch07,Kew06,Kau09}.
The nebular activity detected in the SDSS spectra of early-type
galaxies is, according to standard BPT-diagnostics (see also
\S\ref{sec:SAURONdiagnostic}), generally either LINER-like or
characterised by composite \Hii/Seyfert-like emission (often referred
to as transition objects, or TOs), especially when early-type galaxies
are selected on the basis of their red colour rather than by their
morphology \citep{Sch07}.
Such a kind of nebular emission is most often taken as due to a
central AGN \citep[e.g.,][]{Kau03,Kau09,Kew06}, although unlike for
\Hii\ and Seyfert nuclei there are a number of plausible ionising
mechanisms that can power composite and LINER-like emission.
This is particularly relevant if one considers that the fixed
$3''$-wide aperture of the SDSS spectra usually encompasses large
Kpc-scale regions where diffuse emission such as that observed across
our sample could contribute to the SDSS nebular fluxes.
In fact, other sources than a central AGN may significantly contribute
to power the central LINER-like emission of even much closer galaxies,
and therefore over much smaller physical regions than in the case of
SDSS spectra, typically on 100 pc scales \citep[see, e.g., the case of
  the Palomar survey, ][]{Ho08}.

We can use our previous results on the relative importance of the
various ionising mechanisms at work in and across different kinds of
early-type galaxies in order to better understand the nature of the
nebular emission observed in early-type galaxies in large-scale
surveys, addressing in particular the real extent of AGN activity.
Indeed, we can exploit the integral-field nature of the \sauron\ data
to simulate how our sample galaxies would appear if observed through
wide circular apertures, as is the case for distant early-type
galaxies targeted by the SDSS.
A good starting point for such a comparison between \sauron\ and SDSS
data is the MoSES sample of \citet{Sch07}, which includes
morphologically selected early-type galaxies with redshift values
between $z=0.05$ and $z=0.1$. Using the SDSS pipeline parameter {\tt
  devRad\_r} as a proxy for the effective radius $R_e$, in 68\% of the
MoSES objects the $3''$-wide aperture of the SDSS spectra encompasses
between 32\% and 79\% of $R_e$, an area which is almost always covered
by the \sauron\ observations of our local sample.
At these galactic scales, nebular emission is detected in the MoSES
sample in 18.5\% of the cases, with values for the equivalent width of
the \Oiii\ emission typically (68\%) between 0.8\AA\ and 4.3\AA,
extending most often to 2.9\AA\ and 2.4\AA\ if we consider only
objects with composite or LINER-like emission, respectively. Seyfert
nuclei and star-forming objects indeed make up the majority of the
MoSES galaxies with the most intense emission.

Figure~\ref{fig:SDSSone} shows the radial profiles for the integrated
flux and equivalent width of the \Oiii\ emission over increasingly
large circular apertures, for the 9 \sauron\ galaxies that display
sufficiently intense emission to be detected at the typical
sensitivity of the MoSES spectra.
Out of these 9 objects one is a star-forming galaxy (NGC~3032) whereas
the remaining 8 show integrated values of the \OiiioHb\ ratio in the
log(\OiiioHb)= 0 -- 0.5 range, which is typical of LINER-like emission
(e.g., Ho et al. 1997, TOs can extend to lower \OiiioHb\ values).
These 8 \sauron\ galaxies cover fairly well the range of equivalent
width values observed for most of the MoSES objects with composite and
LINER-like emission (in particular the latter), but fail to extend in
the most intense regime occupied more often by MoSES Seyfert nuclei
and strongly starbursting galaxies.
To some extent this is to be expected considering that Seyfert nuclei
are exceedingly rare, making up only 1.5\% of the MoSES sample, and
given that \Hii-nuclei are most often found in low-mass early-type
galaxies, which are particularly under-represented in the
\sauron\ sample since this is a representative but incomplete sample
of the nearby early-type population \citep{deZ02}. Yet, this may also
suggest that the most intense regime of emission is where truly
nuclear LINER and composite activity can be found in SDSS spectra.
In fact, of the \sauron\ galaxies with detectable LINER-like emission
at the MoSES sensitivity level only half show radio or X-ray signposts
of AGN activity (\S\ref{subsec:AGN}). Furthermore, even in these cases
the normalised profiles for the integrated flux of the \Oiii\ emission
(Fig.~\ref{fig:SDSSone}, top panel) illustrate how the regions
dominated by AGN activity (within $3''$, according to the conservative
estimates of \S\ref{subsec:AGN}) would contribute at most between 30\%
and 55\% of the total \Oiii\ emission detected within the MoSES
apertures.
Thus the \sauron\ galaxies that, if placed at the distance of the
MoSES galaxies, would display the typical values for the equivalent
width of \Oiii\ and for the \OiiioHb\ ratio of the MoSES LINER-like
galaxies are in fact mostly, if not completely, dominated by diffuse
emission powered by other sources other than a central AGN, most
likely post-AGB stars (\S\ref{subsec:OldStars}).

\placefigSDSSone

At first glance, the fraction of \sauron\ objects (9 out of 48, or
$19\pm6$\%) with detectable integrated emission at the MoSES
sensitivity level (hereafter referred simply as ``detectable'') would
also seem consistent with the fraction of the MoSES sample displaying
nebular emission (18.5\%), even when the fraction of star-forming
systems (1 out 48, or $2\pm2$\% compared to 4.3\%) or of galaxies with
composite or LINER-like emission (8 out 48, or $17\pm5$\% compared to
12.6\%) are considered separately.
Yet, it should be noted that the objects with composite emission in
the MoSES sample come typically (68\%) with log(\OiiioHb) values
between -0.25 and 0.28, whereas the 8 ``detectable'' and
non-starforming \sauron\ galaxies show log(\OiiioHb) values between
0.29 and 0.43, a range that is much closer to that of the MoSES
galaxies with LINER-like emission (where log(\OiiioHb) spans between
0.18 and 0.55).
To put it in a more direct way, the fraction of the MoSES sample that
displays log(\OiiioHb) values in the 0 -- 0.5 range, which is
generally associated with LINER-like emission, is just 8.3\%, or only
about half the fraction of the ``detectable'' \sauron\ galaxies with
integrated log(\OiiioHb) values in the same range.
It is possible that the lack of ``detectable'' \sauron\ galaxies with
integrated values for \OiiioHb\ ratio that are more typical of
composite activity is due to small-number statistics and the
incomplete character of the \sauron\ sample. Indeed, as for the
objects displaying star formation and Seyfert activity, also the MoSES
galaxies with composite emission live preferentially in low mass
early-type galaxies, which are particularly under-represented in the
\sauron\ sample.
Alternatively, it is possible that some of the MoSES galaxies would
not be classified as early-type if observed at a much closer distance.
At least for a local point of view, a more quantitative analysis will
have to wait for a complete integral-field survey of the nearby
early-type galaxy population, which will be possible once the data
from the ATLAS$^{\rm 3D}$ 
campaign\footnote{See also {\tt http://purl.org/atlas3d}}
are at hand.

Notwithstanding the limitations of the present analysis, the picture
we have drawn for SDSS galaxies with \OiiioHb\ ratios typical of
LINER-like emission (of which we have an overabundance of
``detectable'' \sauron\ examples) would reinforce the recent results
of \citet{Sta08}, who advocate the importance of photo-ionisation by
post-AGB stars in "retired" galaxies to explain the LINER population
in the SDSS survey.
Yet, we note here that 2 out of the 8 ``detectable'' \sauron\ objects
with LINER-like values of the \OiiioHb\ ratio are most likely powered
by post-AGB associated to a recent star-formation episode (NGC~3156
and NGC~3489, \S\ref{subsec:HIERs}) rather than to the older bulge
population.
This kind of sources could be particularly common in the SDSS LINER
population, considering that such \sauron\ post-starbursting systems
appear to be more easily ``detectable'' in SDSS spectra (50\%, or 2
out of 4) than is the case of the other \sauron\ galaxies with diffuse
LINER-like emission but uniformly old stellar populations (25\%, or 6
out 25).
In fact, such a post-starburst LINER-like activity could contribute to
explain the findings of \citet{Gra07}, that red-sequence SDSS galaxies
with strong LINER-like emission show younger stellar populations, in
particular in less-massive systems, such as NGC~3156 and NGC~3489.\\

To conclude, our \sauron\ integral-field data suggest that in very
few, if any, of the SDSS galaxies which display only modest values for
the equivalent width of the \Oiii\ line (less than $\sim$2.4\AA) and
LINER-like values for the \OiiioHb\ ratio, the nebular emission is
truly powered by AGN activity. Only the most intense, and rarer,
manifestations of LINER nuclear activity can be detected against both
the stellar background encompassed by the SDSS aperture and the
diffuse emission observed in early-type galaxies, unless to consider a
very small minority of the closest SDSS objects.
For SDSS samples that share these equivalent-width and line-ratio
characteristics, the \Oiii$\lambda5007$ line is thus problematic as
a diagnostic of low levels of black-hole growth, and LINERs may
represent a mixed population that does not uniformly trace
low-Eddington ratio accretion.

\section{Conclusions}
\label{sec:conclusions}

Building on our previous investigation of the incidence, morphology
and kinematics of the ionised gas in the early-type galaxies of the
\sauron\ sample, in this paper we set out to address the question of
what is powering the nebular emission observed in them.

To constrain the possible sources of gas excitation we resorted to a
variety of ancillary data ranging from radio to X-ray wavelengths,
drew from complementary information on the gas kinematics, stellar
populations and galactic potential from our \sauron\ data, and
introduced a \sauron-specific diagnostic diagram based on the
\NioHb\ line ratio as a gauge for the hardness of the ionising
continuum.
Using SDSS data for galaxies with strong emission we have shown that
the \OiiioHb\ vs \NioHb\ diagram is a powerful tool to isolate
star-forming regions and to clearly separate Seyfert and LINER nuclear
activity.
To best use the \NioHb\ diagnostic we re-extracted more reliable
\Ni\ measurements from the \sauron\ data using a new array of model
and empirical templates based on the MILES stellar library. Such
empirical templates come from fits to high-quality \sauron\ spectra
extracted over regions in most cases devoid of emission lines and with
values for the absorption-line strengths that cover well the range
observed across our sample.

The considerable range of values for the \OiiioHb\ line ratio found
both across the \sauron\ sample and within single galaxies has
prompted us to explore the relative importance of a central AGN, star
formation, old UV-bright stars, shocks and of the interaction with the
hot phase of the ISM in powering the nebular emission from early-type
galaxies.
The role of these ionisation sources can be summarised as follow:

\begin{itemize}

\item[]{\it pAGB Stars.\/} A tight correlation between the stellar
  surface brightness and the flux of the \Hb\ recombination line
  throughout the vast majority of our sample galaxies points to a
  diffuse and old stellar source as the main contributor of ionising
  photons in early-type galaxies.  Based on simple ionisation-balance
  arguments we have shown that since the initial suggestion of
  \citet{Bin94} pAGB stars remain the favourite candidate for powering
  the ionised-gas emission of early-type galaxies, which urges further
  investigations into the present discrepancy between observations and
  models as regards their total numbers.

\item[]{\it AGN.\/} The presence of a radio or X-ray core suggests the
  additional role of AGN photoionisation in approximately 30\% of the
  galaxies in our sample, although the radial profile of the
  \Hb\ recombination line shows that a central source can be
  responsible for the observed emission at best only within a few
  hundred parsecs from the centre, corresponding to the innermost
  $2''-3''$.

\item[]{\it Shocks.\/} Fast shocks are unlikely to be an important
  source of ionisation for the extended nebular emission observed in
  early-type galaxies, even where spiral or integral-sign gas
  structures are accompanied by a similar \OiiioHb\ feature. Indeed,
  the shock velocities required to power the nebular emission can
  hardly be attained in the potential of our sample galaxies, in
  particular given the relatively relaxed gas kinematics that is
  observed in them.

\item[]{\it OB-stars.\/} Extremely low values of the \OiiioHb\ ratio
  leave no doubt that OB-stars are powering most of the nebular
  emission observed in 6\% of the \sauron\ sample galaxies (3/48),
  whereas the presence of regular dust morphologies, relaxed gas
  kinematics, young stellar populations, molecular gas, PAH features
  and strong FIR fluxes further support the case for on-going star
  formation in up to 10\% of the sample (5/48).
  In these objects star formation appears confined mostly to
  circumnuclear regions, which in two cases is confirmed by
  \NioHb\ diagnostic diagrams that highlight the role of other sources
  of ionisation outside of the ring regions, for instance of a central
  AGN towards the centre.
  Yet, star formation may have proceeded also outside these
  circumnuclear regions, in particular in the least massive galaxies
  like NGC~3032 where younger stellar populations are found throughout
  the entire \sauron\ field of view.

\item[]{\it Post-starburst pAGB Stars.\/} For another 10\% of the
  \sauron\ sample, consisting too of low-mass objects, the intense and
  highly-ionised emission relates also to a recent and spatially
  extended star-formation episode, although the principal ionising
  sources are no longer OB-stars but the pAGB population associated to
  the younger stellar subcomponent.

\item[]{\it Interaction with the Hot Phase.\/} At the opposite end of
  the mass spectrum, we find that in the brightest and slowly- or
  non-rotating early-type galaxies the ability to retain massive halo
  of hot, X-ray emitting gas corresponds to a distinct low-ionisation
  character of the emission-line regions and to a remarkable spatial
  coincidence between optical and X-ray emitting features.
  These findings testify to an interaction between the hot and warm
  phases of the ISM that provides an additional excitation mechanism
  besides photoionisation by old stellar sources.
 
 \end{itemize}

The prominent role of old stellar sources over other ionising
mechanisms in powering the nebular emission of early-type galaxies,
taken together with the extended character that ionised gas always
displays in our sample, motivated us to explore the nature of the
ionised-gas emission that is observed in early-type galaxies at larger
distances through large physical apertures, as in the case of the SDSS
spectroscopic data.
We have exploited the integral-field nature of our data to integrate
the \sauron\ spectra of our sample galaxies over increasingly wider
circular apertures as to mimic the spectra of increasingly distant
objects observed through the fixed 3$"$-wide SDSS aperture. We
measured the values for the \Oiii\ equivalent width and the
\OiiioHb\ ratio for the total emission in such apertures and compared
our results with the emission-line measurements for the
morphologically-selected sample of early-type galaxies of
\citet{Sch07}.
Even accounting for the incomplete character of the \sauron\ sample,
our data suggest that in very few, if any, of the SDSS galaxies which
display only modest values for the equivalent width of the \Oiii\ line
(less than $\sim $2.4\AA) and LINER-like values for the
\OiiioHb\ ratio, the nebular emission is truly powered by AGN
activity. Only the most intense, and rarer, manifestations of LINER
nuclear activity can be detected against both the stellar background
encompassed by the SDSS aperture and the diffuse emission of
early-type galaxies. The latter, when detected through SDSS spectra,
is usually powered by pAGB stars associated either to old stellar
populations or to a recent star-formation episode.

\section*{Acknowledgements}
MS is grateful to Mark Allen, Aaron Barth, Igor Chilingarian, Alison
Crocker, Brent Groves, Guenevieve Graves, Luis Ho, Sadegh Khochfar,
John Magorrian, Gary Mamon, Ralph Napitowski, Carlo Nipoti, Martin
Hardcastle and Philipp Podsiadlowski for many fruitful discussions. We
are also grateful to Alexandre Vazdekis and Mark Allen for providing
us with their latest models prior to publication, and to the anonymous
referee for his/her useful comments. MS and MC also acknowledge
support from their respective STFC Advanced Fellowships.

%

\label{lastpage}
\end{document}

%% file: my_commands.tex
\newcommand{\kms}{\mbox{${\rm km\;s^{-1}}$}}
\newcommand{\cms}{\mbox{${\rm cm\;s^{-1}}$}}
\newcommand{\pccm}{\mbox{${\rm cm^{-3}}$}}
\newcommand{\sauron}{{\texttt {SAURON}}}

\newcommand{\Oiii}{[{\sc O$\,$iii}]}

\newcommand{\Oi}{[{\sc O$\,$i}]}
\newcommand{\Ha}{H$\alpha$}
\newcommand{\Hb}{H$\beta$}
\newcommand{\Hi}{{\sc H$\,$i}}
\newcommand{\Hii}{{\sc H$\,$ii}}
\newcommand{\Ni}{[{\sc N$\,$i}]}
\newcommand{\Nii}{[{\sc N$\,$ii}]}
\newcommand{\Sii}{[{\sc S$\,$ii}]}
\newcommand{\OiiioHb}{\Oiii/\Hb}
\newcommand{\OioHa}{\Oi/\Ha}

\newcommand{\NioHb}{\Ni/\Hb}

\newcommand{\Sg}{$\sigma_{\rm gas}$}
\newcommand{\eg}{e.g.,}
\newcommand{\ie}{i.e.,}
\newcommand{\apj}{ApJ}

\newcommand{\mnras}{MNRAS}

%% file: figures_commands.tex
\newcommand{\placefigNioHbone}{
\begin{figure*}
\begin{center}
\includegraphics[width=0.9\textwidth]{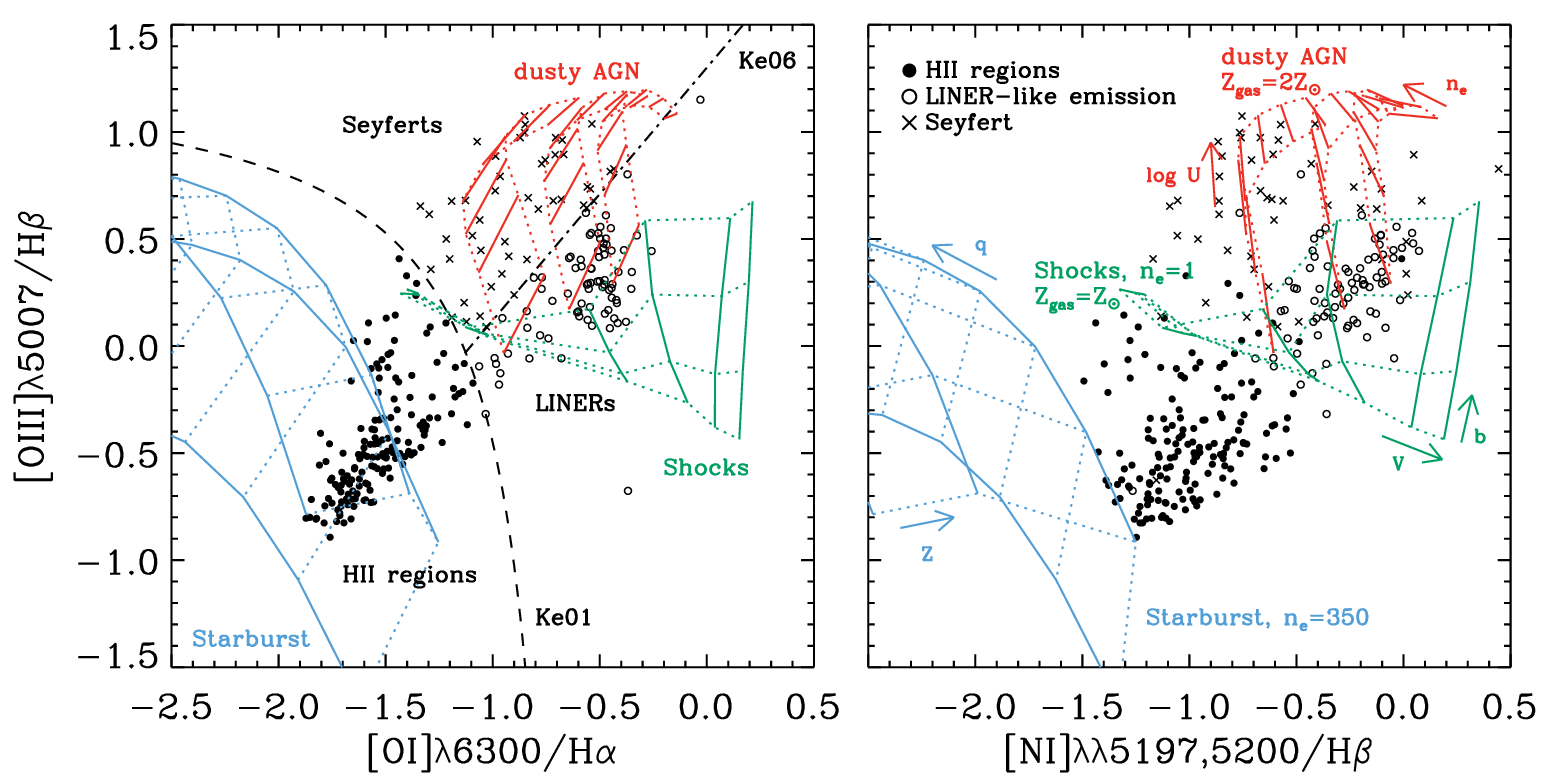}
\end{center}
\caption[]{
The standard \Oi/\Ha\ vs. \Oiii/\Hb\ BPT diagram of Veilleux \&
Osterbrock (left) compared to the
\sauron\ \Ni/\Hb\ vs. \Oiii/\Hb\ diagnostic (right). In both panels
the red, blue and green lines show the predictions of the MAPPINGS-III
models for gas that is photoionised by a central AGN, by O-stars, or
that is excited by shocks, respectively.
The AGN grids are from the dusty, radiation pressure-dominated models
of \citet{Gro04b} and adopt three values for the index $\alpha$ of the
power-law AGN continuum $f_\nu\propto\nu^{\alpha}$
($\alpha=-1.7,-1.4,-1.2$ from left to right). In each AGN model grid
the solid lines trace the dimensionless ionisation parameter $\log{U}$
(defined as $\log{q/c}$), which increases with the \OiiioHb\ ratio
from $\log{U}=-3.0,-2.6,-2.3,-2.0,-1.6,-1.3,-1.0$, whereas the dotted
lines show the adopted values for the electron density of $N_e=10^2$
and $10^4\,\pccm$, with smaller values of the \NioHb\ ratio
corresponding to larger $N_e$ values.
The starburst grids are from \citet{Dop00}, and assume a gas density
of $N_e=350\,\pccm$ and use a spectral energy distribution obtained
from Starburst99 \citep{Lei99} models for an instantaneous
star-formation episode. The grids assume a range of metallicity $Z$
for both stars and gas in the starburst, shown by the solid lines for
$Z=0.2, 0.4, 1.0, 2.0\,{\rm Z}_\odot$, and different values of the
ionising parameter $q$, shown by the dotted lines for $q=0.5, 1, 2, 4,
8, 15, 30 \times 10^7\,\cms$.
Similarly, for the shock grids (without precursor \Hii\ region), the
solid lines show models with increasing shock velocity $V_{s}=150,
200, 300, 500, 750, 1000\,\kms$, and the dotted lines models with
magnetic parameter $b=0.5, 1.0, 2.0, 4.0$. The shock grids are from
\citet{All08} and assume an electron density of $N_e=1\,\pccm$ and a
solar value for the gas metallicity.
In both panels the filled circles, the crosses and the open circles
show the values of the emission-line ratios observed in the central
3$''$ of 284 galaxies with SDSS spectra, where according to the limits
draw by \citet{Kew01} and \citet{Kew06} in the \OioHa\ diagram the
nebular emission is most likely arising from \Hii\ regions, from a
Seyfert AGN or from LINER-like emission regions, respectively.}
\label{fig:NIone}
\end{figure*}
}

\newcommand{\placefigMILESone}{
\begin{figure}
\begin{center}
\includegraphics[width=\columnwidth]{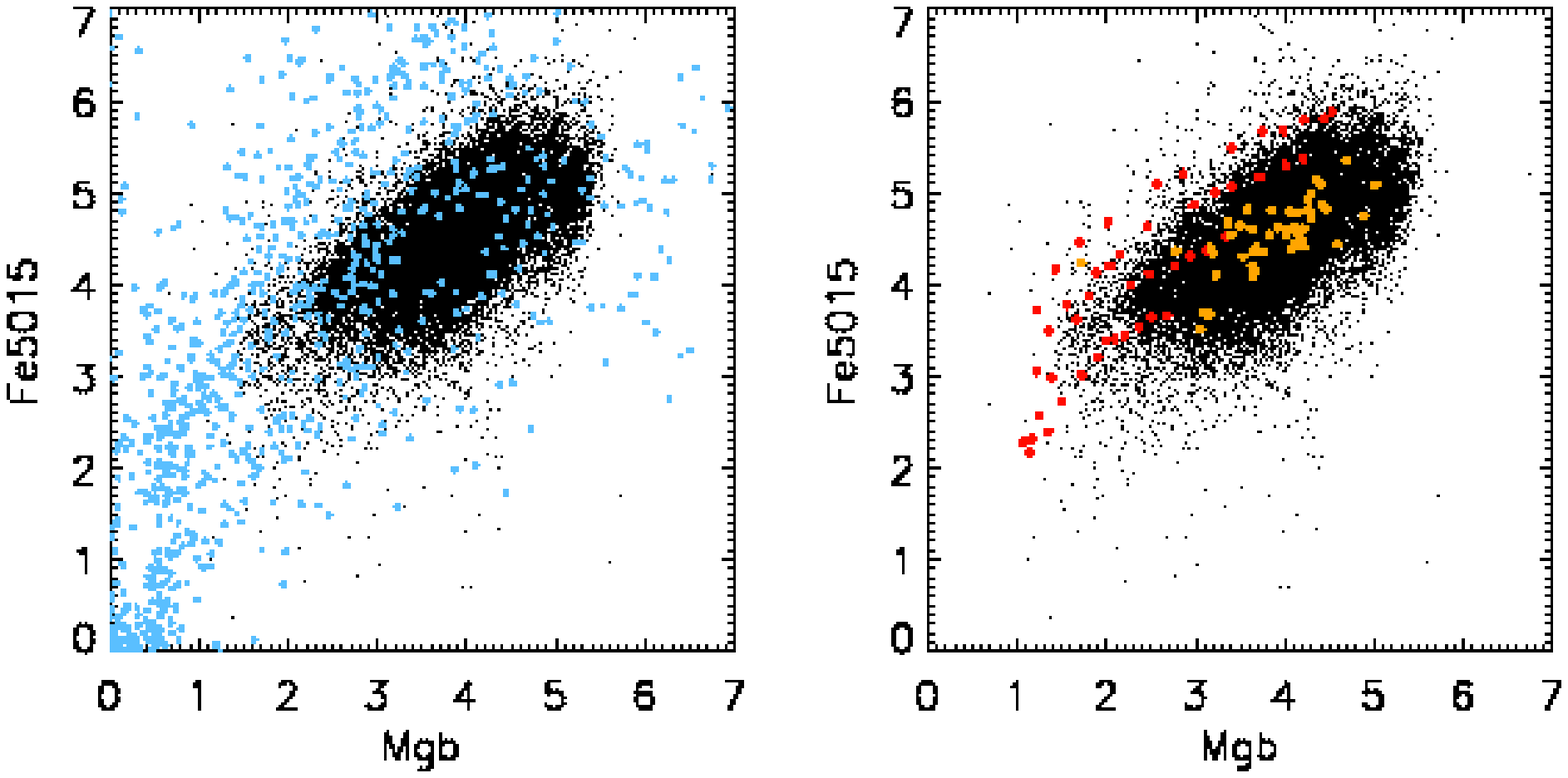}
\end{center}
\caption[]{
Distribution of the values for the Lick Mg$b$ and Fe5015 indices
observed across the \sauron\ sample (from the maps of Paper~VI, black
dots) compared to that observed across the MILES stellar library
(left, blue circles) or for the single-age models of Vazdekis et
al. and the empirical templates based on matching real
\sauron\ spectra (right, red and orange circles, respectively). 
The MILES stars provide a poor coverage of the region in the Mg$b$
vs. Fe5015 plane that is most populated by the \sauron\ data, which
causes the continuum fit process to adopt unphysical combinations of
stars and thus leads to spurious emission-line measurements, in
particular to exceedingly large \Ni\ fluxes. This is not the case when
simple stellar population models and empirical templates are adopted,
although the quality of the fit is formally not as good.}
\label{fig:MILESone}
\end{figure}
}

\newcommand{\placefigAGNone}{
\begin{figure*}
\begin{center}
\includegraphics[width=\textwidth]{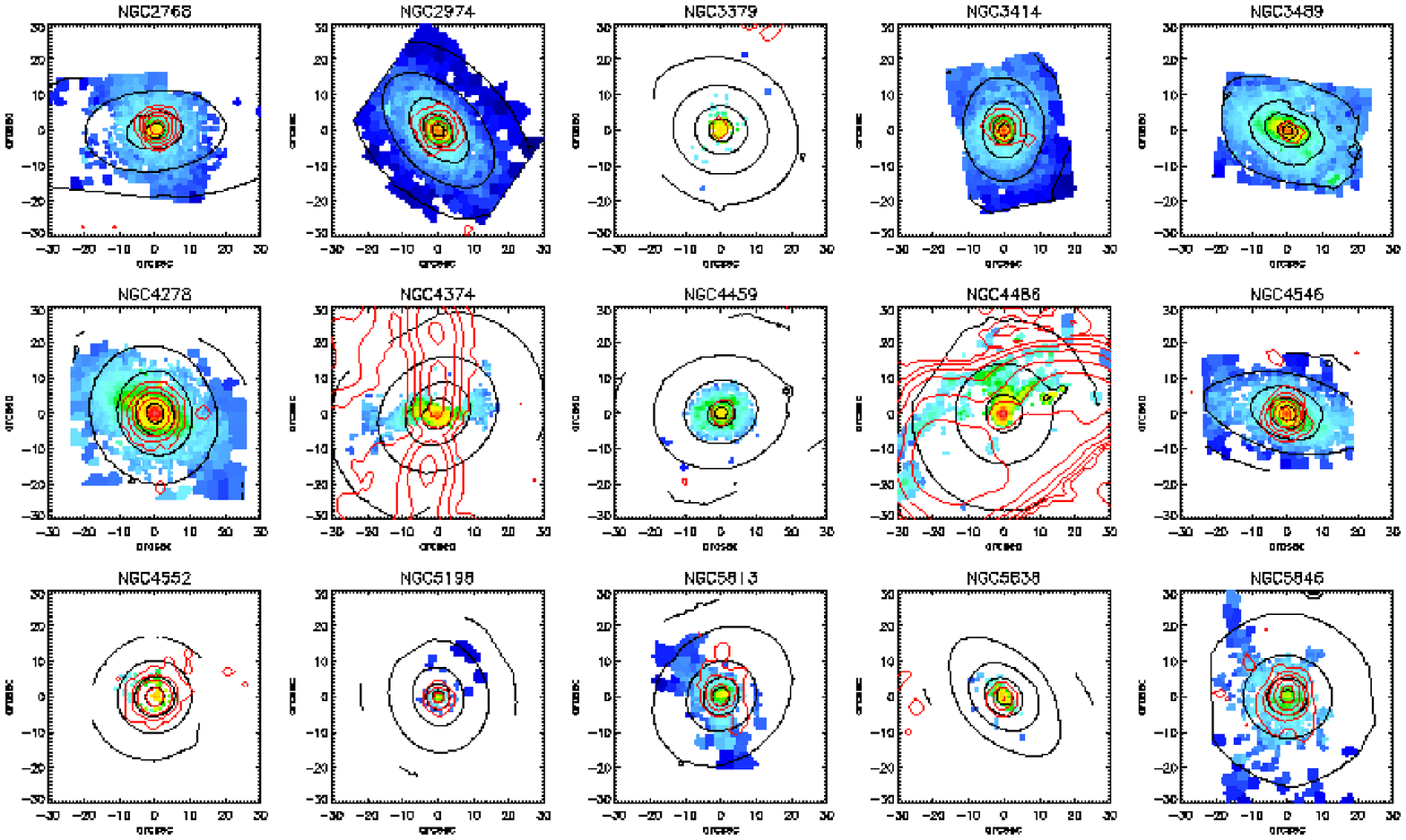}
\end{center}
\caption[]{
\sauron\ \Hb\ emission vs. VLA measurements for the 1.4~Ghz radio
continuum from the FIRST survey and for galaxies with compact radio
cores. As in Paper~V the \Hb\ fluxes are colour-coded in a logarithmic
scale with values ranging between $10^{-14}$ and $10^{-18} \rm
erg\,s^{-1}cm^{-2} arcsec^{-2}$. The red contours show the 1.4~Ghz
fluxes also in logarithmic scale, starting from a level of $0.45\,\rm
mJy/beam$ (corresponding to 3 times the typical RMS of the FIRST
images) and doubling with each contour, except for NGC~4278, NGC~4374,
NGC~4486 and NGC~4552 for which the 1.4~Ghz fluxes changes by a factor
5 between each line. The beam of the FIRST images is circular and has
a FWHM of 5\farcs4.}
\label{fig:AGNone}
\end{figure*}
}

\newcommand{\placefigAGNtwo}{
\begin{figure*}
\begin{center}
\includegraphics[width=\textwidth]{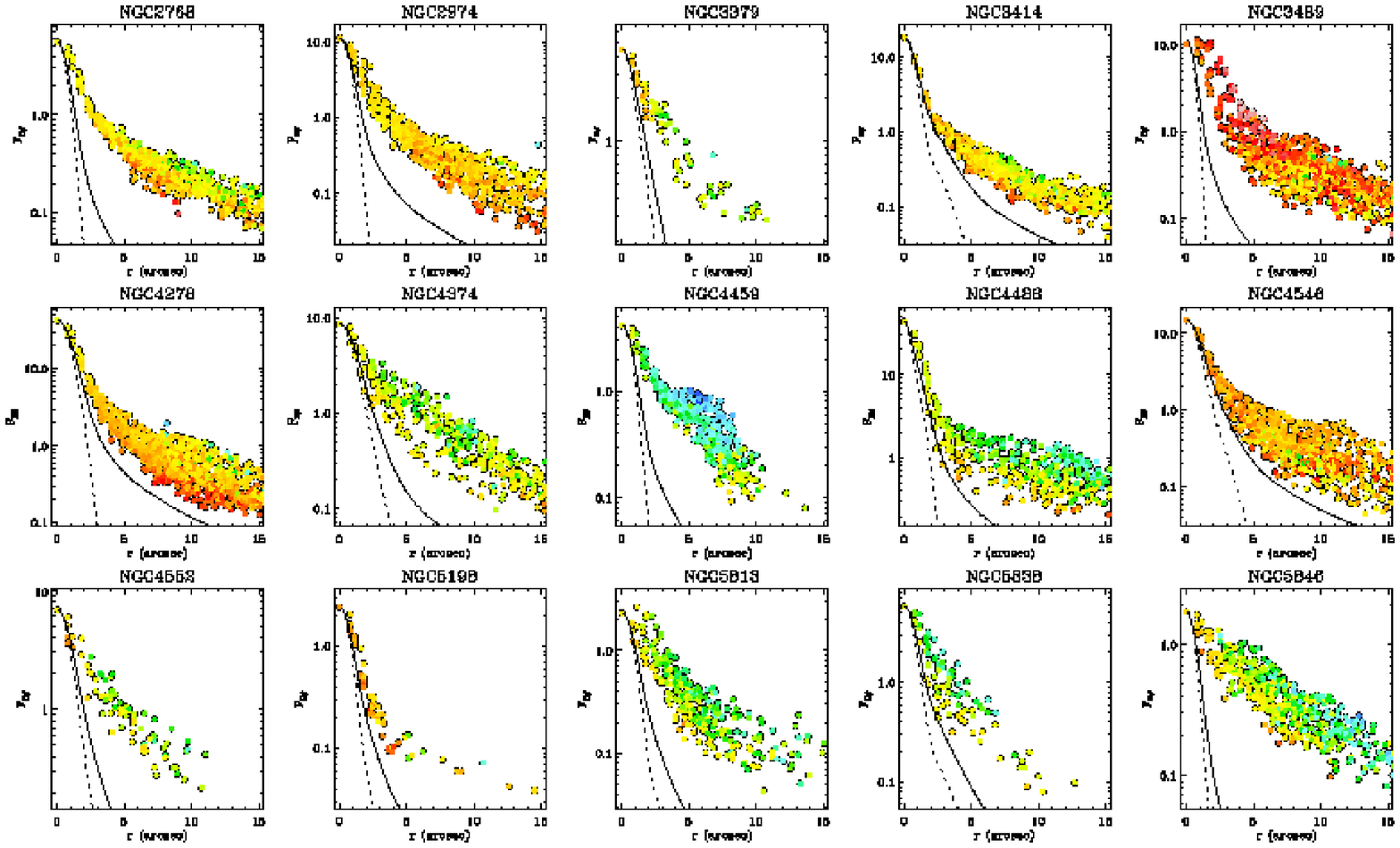}
\end{center}
\caption[]{
Radial profiles for the \Hb\ flux values in galaxies with 1.4~Ghz
radio cores. The \Hb\ fluxes are in units of $10^{-16} \rm
erg\,s^{-1}cm^{-2} arcsec^{-2}$ and are plotted only out to 15$''$.
In each panel the solid line shows the $r^{-2}$ radial decrement that
is expected in the case AGN photoionisation and under the special
conditions of constant gas density and filling factor, accounting for
the impact of atmospheric blurring. The point-spread function of the
\sauron\ observations, from Paper~III, is shown by the dotted
lines. Both profiles have been normalised to the central
\Hb\ emission. The \Hb\ data are colour-coded according to the
logarithm of the \Oiii/\Hb\ ratio, going from blue to green and red
for values of -1, 0 and 1.}
\label{fig:AGNtwo}
\end{figure*}
}

\newcommand{\placefigAGNthree}{
\begin{figure*}
\begin{center}
\includegraphics[width=0.9\textwidth]{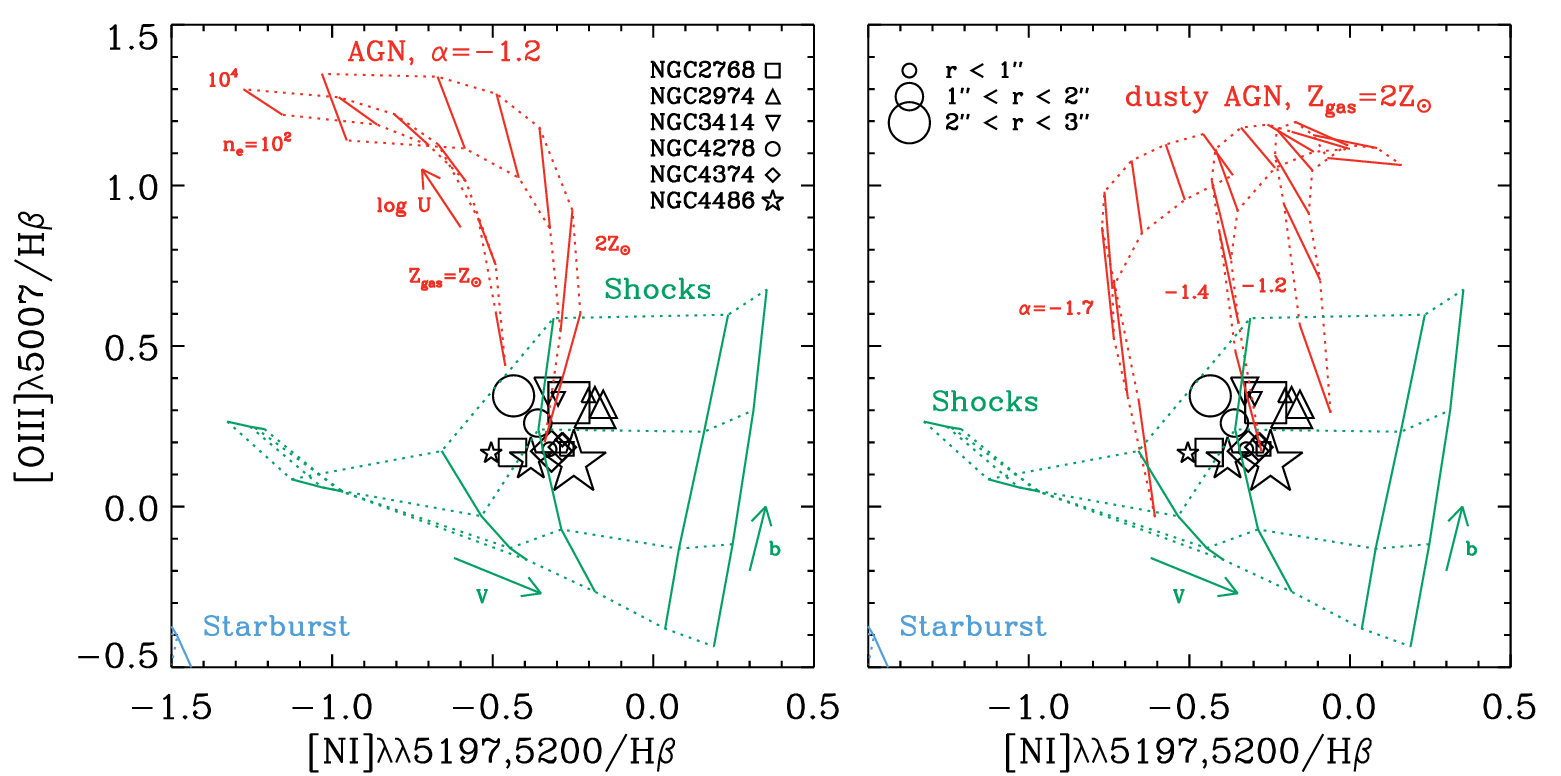}
\end{center}
\caption[]{
\sauron\ \Ni/\Hb\ vs. \Oiii/\Hb\ diagnostic diagram for galaxies with
1.4~Ghz radio cores. The different symbols correspond to each of the
galaxies with detected \Ni\ emission, and the size of the symbols
indicate the radial range over which the average values of the
\Ni/\Hb\ and \Oiii/\Hb\ ratios have been computed.
The colored grids show the same MAPPINGS-III models as in
Fig.\ref{fig:NIone}, except for the AGN photoionisation models shown
in the left panel, which do not include dust, correspond to a single
value of the power-law index of the AGN continuum of $\alpha=-1.2$ and
are shown also for solar values of the gas metallicity. The range of
the dimensionless ionising parameter of the dust-free AGN models is
also somewhat different, including values of
$\log{U}=-3.3,-3.0,-2.6,-2.3,-2.0,-1.6$.}
\label{fig:AGNthree}
\end{figure*}
}

\newcommand{\placefigSFone}{
\begin{figure*}
\begin{center}
\includegraphics[width=0.9\textwidth]{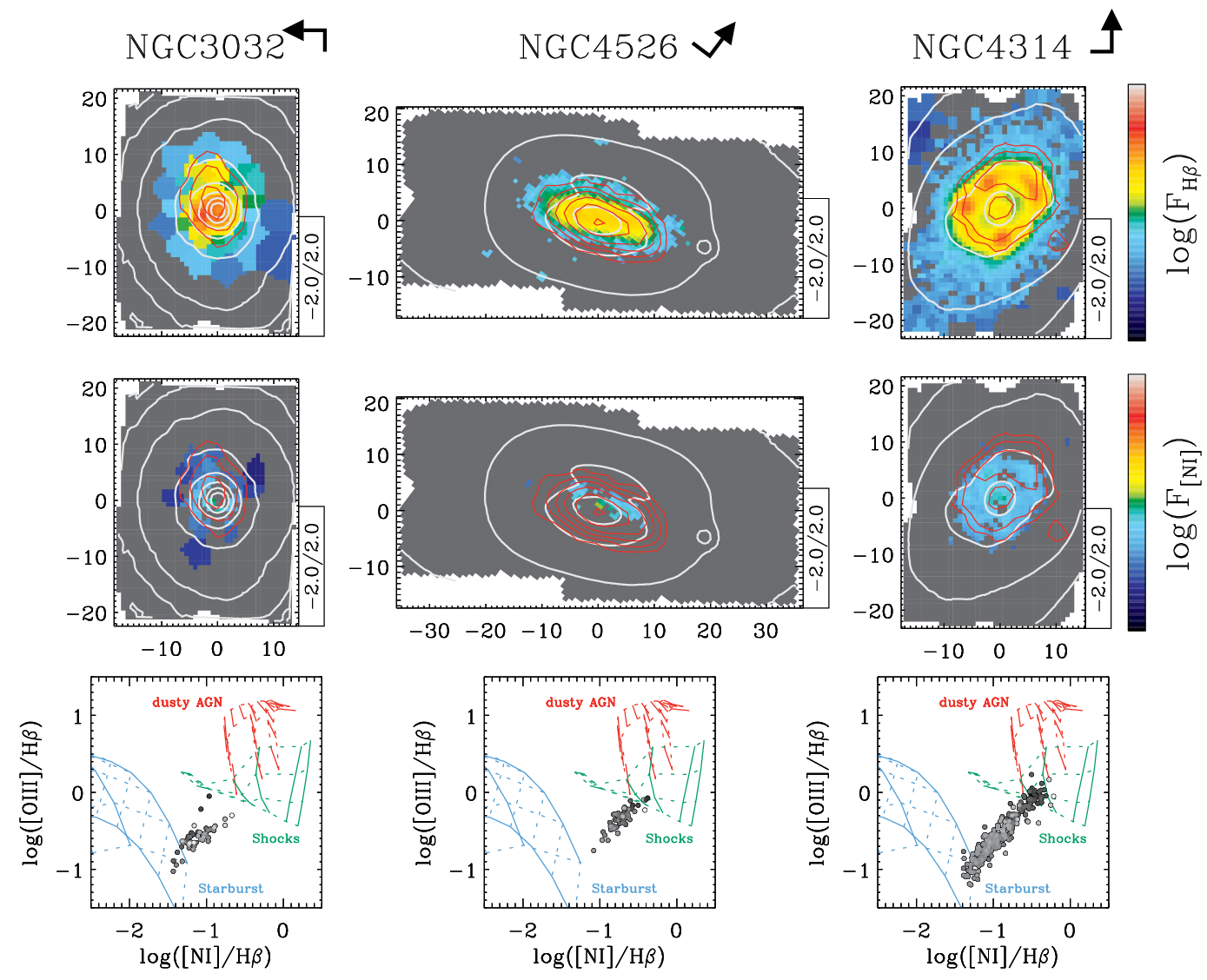}
\end{center}
\caption[]{
Two examples of lenticular galaxies with extremely regular dusty disks
and gas kinematics where star formation is on-going, shown together to
the Sa galaxy NGC~4314 which hosts a well-known circumnuclear
star-forming ring.
The top and middle panels show maps for the \Hb\ and
\Ni$\lambda\lambda5198,5200$ emission. Superimposed to these maps are
1.4~Ghz radio continuum contours ({\it red lines}) from the FIRST-VLA
images illustrating the resolved character of the synchrotron and
free-free emission.
As in the Paper~V and for the \sauron\ maps in following figures, the
white contours follow the stellar surface brightness, the scale is in
arcsec and the side box show the range of the plotted quantity, in
this case the flux of the \Hb\ and \Ni\ lines, in logarithmic scale.
The \Ni\ doublet is detected only where \Oiii\ and \Hb\ emission is
the strongest, which allows to trace these regions in
\NioHb\ diagnostic diagrams shown in the lower panels. Similar to the
case of the most intense starburst observed in SDSS data
(Fig.~\ref{fig:NIone}), all three galaxies share a distinct diagonal
trend in the \NioHb\ diagnostic, with the central and outermost
regions (shown in darker and lighter points, respectively) departing
the most from the starburst model grids and falling closer to the AGN
or shock grids.}
\label{fig:SFone}
\end{figure*}
}

\newcommand{\placefigDiffuseone}{
\begin{figure*}
\begin{center}
\includegraphics[width=0.9\textwidth]{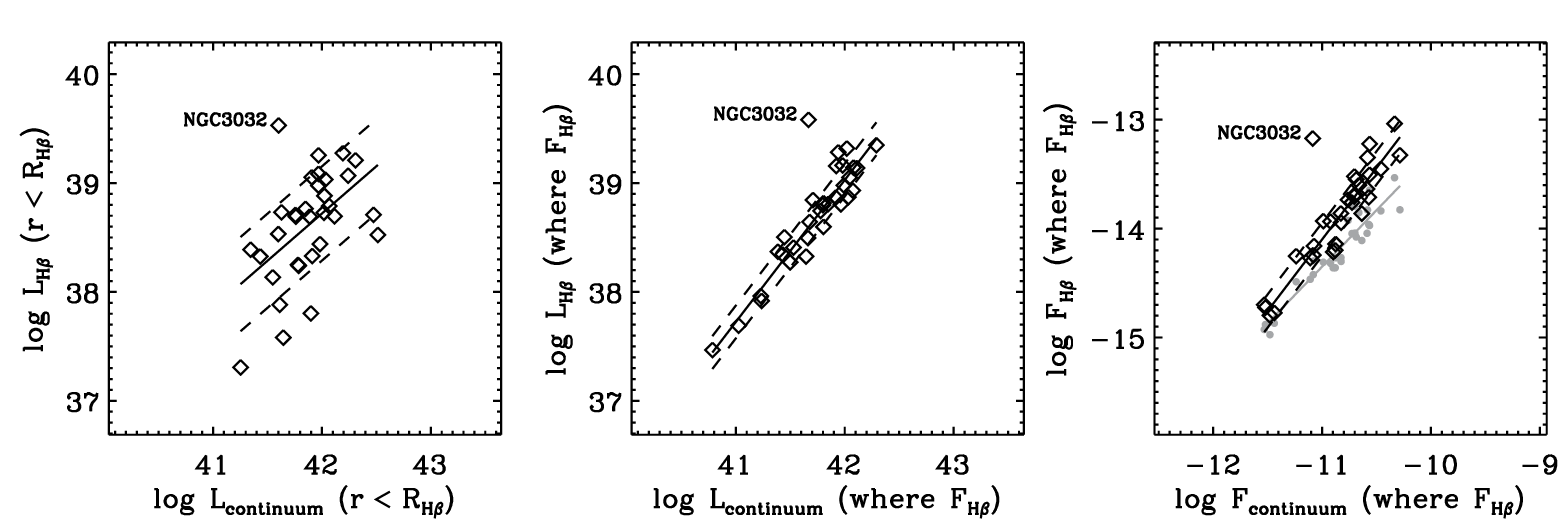}
\end{center}
\caption[]{
The connection between the integrated luminosity of the nebular
emission and that of the stellar continuum in early-type galaxies.
Left: Correlation between the luminosity of the stellar continuum
$L_{\rm cont}$ and that of the \Hb\ emission $L_{\rm H\beta}$ observed
in the \sauron\ spectra when both quantities are integrated within the
average radius of the emission-line regions detected in each of our
sample galaxies. The solid and dashed line show the best fitting
power-law fit to the observed distribution and associated $1\sigma$
scatter, respectively.
Middle: Same as left but now the total \Hb\ luminosity is juxtaposed
to the stellar luminosity integrated only where the nebular
\Hb\ emission is detected. A much tighter correlation emerges in this
case.
Right: Same as {\t middle} but now showing the correlation between the
integrated nebular and stellar fluxes. A tight correlation persists,
indicating that the $L_{\rm cont}$--$L_{\rm H\beta}$ relation is not
driven by the distance distribution of our sample galaxies. The light
grey points show the integrated \Hb\ detection limits in each
galaxy. Their position and alignment differs significantly from that
of our integrated \Hb\ fluxes, dispelling also any doubts that the
observed correlations are due to the limited range of our equivalent
width sensitivity.
In all panels the position of NGC~3032, our most significant outlier,
is indicated, and can be explained in terms of intense star-formation
activity.}
\label{fig:Diffuseone}
\end{figure*}
}

\newcommand{\placefigDiffusetwo}{
\begin{figure*}
\begin{center}
\includegraphics[width=0.9\textwidth]{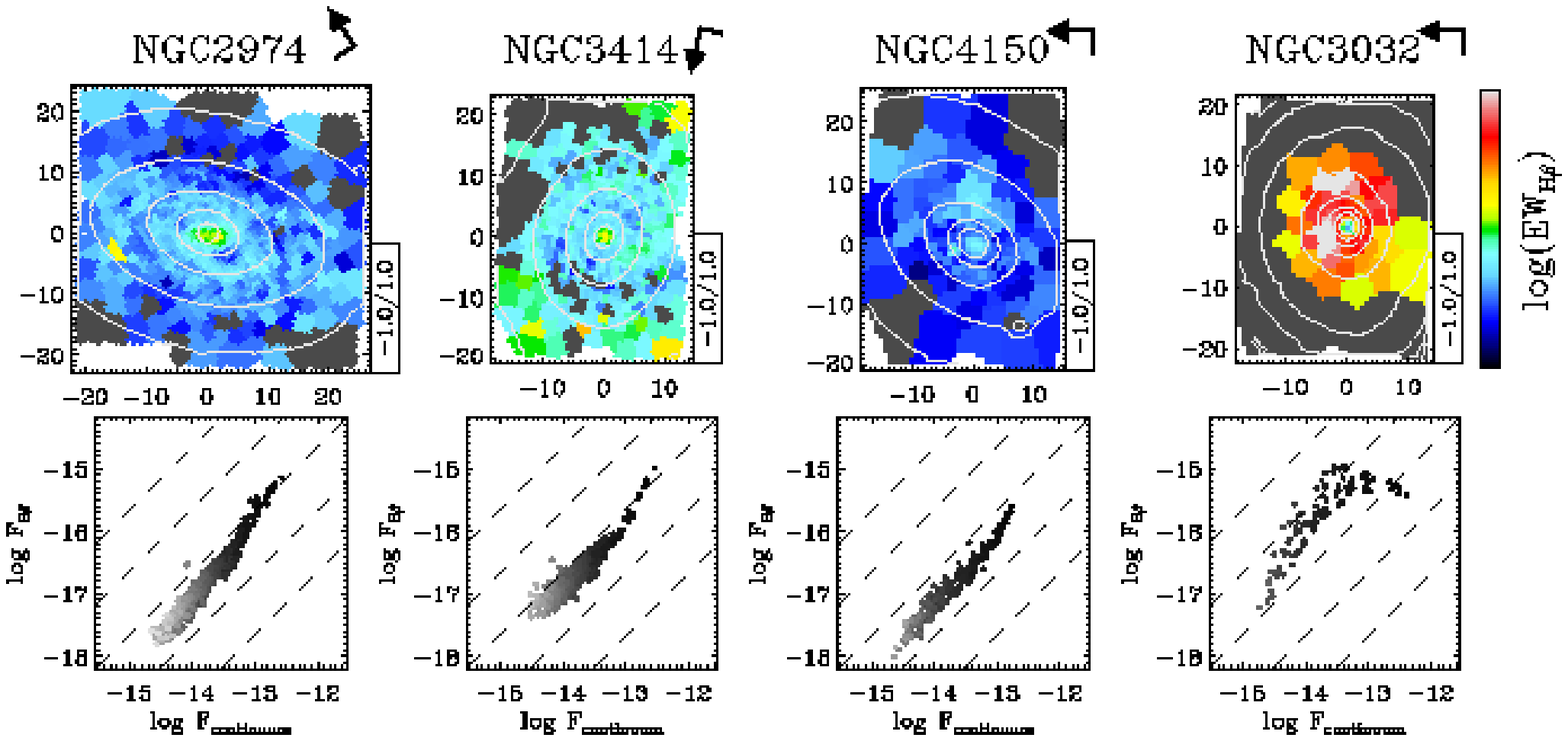}
\end{center}
\caption[]{
Examples of the local connection between the flux of the nebular
emission and stellar surface brightness in early-type galaxies. Maps
for the equivalent width of the \Hb\ recombination line ({\it top
  panels}) display remarkably similar values for this quantity across
most of the \sauron\ field of view, reflecting how closely the flux of
the \Hb\ line follows that in the stellar continuum ({\it lower
  panels}).
Such a correlation breaks down in the central regions of our sample
galaxies (shown with darker points in the lower panels) where AGN
photoionisation can be important (\S\ref{subsec:AGN}), or in objects
where OB-stars are the main source of ionisation (e.g. in NGC~3032,
rightmost panels)}
\label{fig:Diffusetwo}
\end{figure*}
}

\newcommand{\placefigDiffusethree}{
\begin{figure}
\begin{center}
\includegraphics[width=\columnwidth]{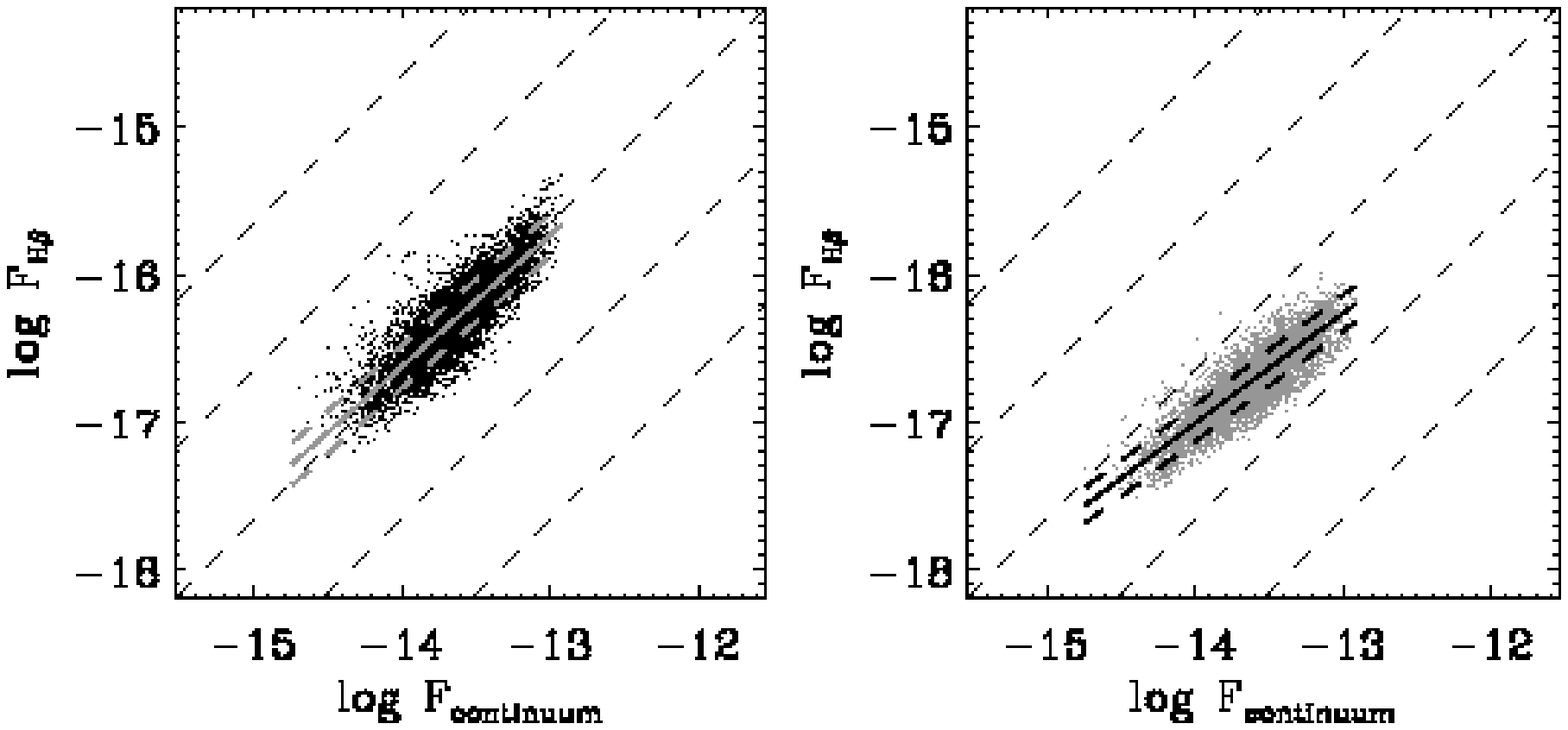}
\end{center}
\caption[]{
Correlation between the flux of the stellar continuum and the flux of
the \Hb\ line (left panel) observed in each of the bin of our sample
galaxies where emission is detected, except for the central 3" and
excluding the objects with obvious star formation.
For each of the plotted galaxies, the continuum and nebular flux
values have been rescaled to the same average value, of
$2.9\times10^{-14}$ and $6.4\times10^{-17} {\rm erg\, s^{-1}
  cm^{-2}}$, respectively.
The right panel shows the same correlation, but for the detection
limits on the \Hb\ line, normalised by the same factor used for the
observed \Hb\ values.}
\label{fig:Diffusethree}
\end{figure}
}

\newcommand{\placefigDiffusefour}{
\begin{figure}
\begin{center}
\includegraphics[width=\columnwidth]{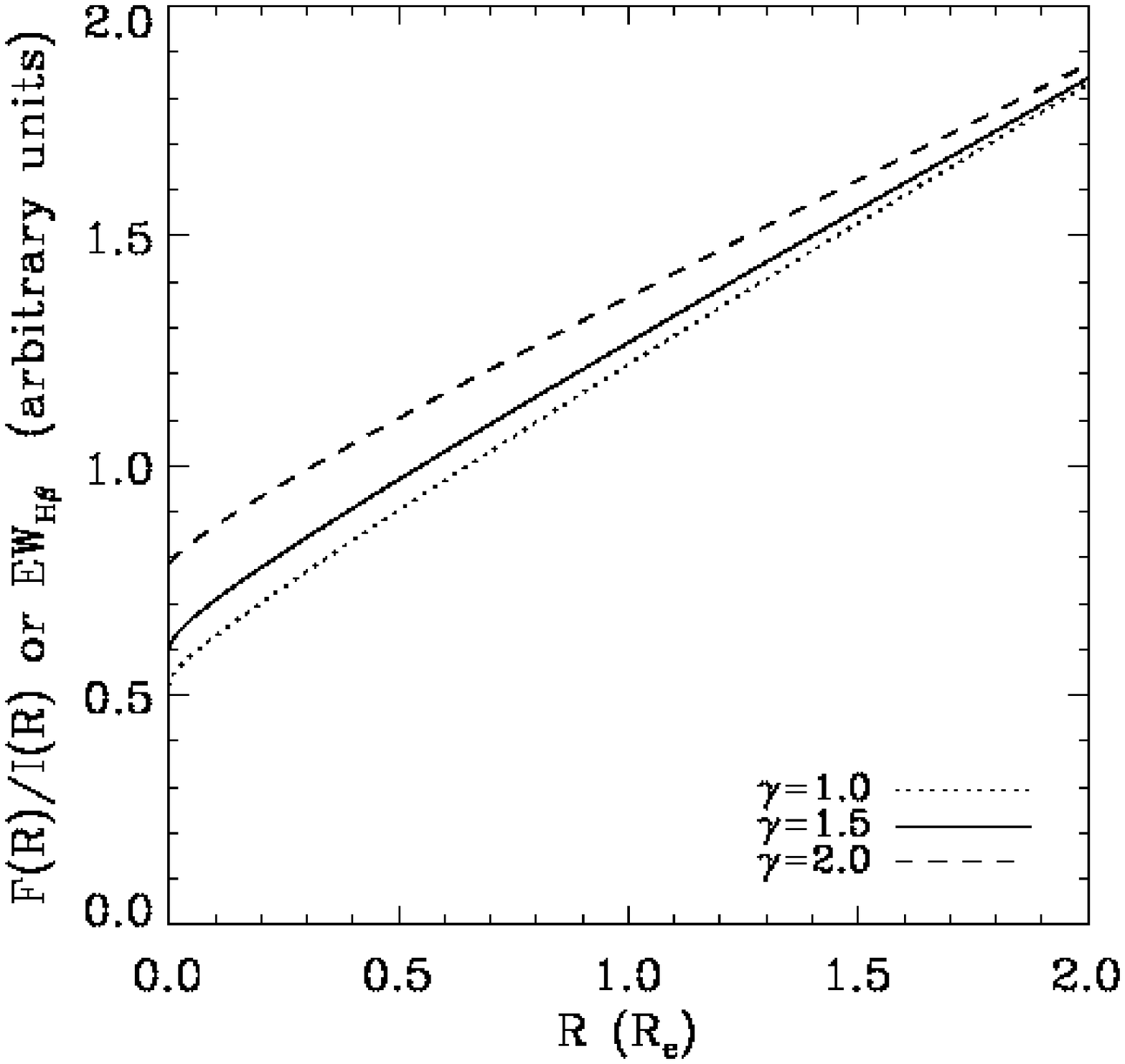}
\end{center}
\caption[]{
Predicted radial profile for ratio of the ionising radiation $F(R)$
from evolved stellar sources diffuse throughout the galaxy to the
surface stellar brightness $I(R)$. Assuming that a constant fraction
of $F(R)$ is reprocessed as \Hb\ emission, the plotted $F/I$ profile
provides a prediction for the radial variation for the equivalent
width of \Hb. The model further assumes a spherical geometry and three
different kind of $\gamma$--models for the intrinsic luminosity
density $L(r)$, 
approximating a \citet[][$\gamma=1.5$]{deV48} or
corresponding to \citet[][$\gamma=1$]{Her90}, and
\citet[][$\gamma=2$]{Jaf83} surface-brightness profiles. Independent
of the chosen $L(r)$, the equivalent width of the recombination lines
should increase towards the outskirts of early-type galaxies according
to this simple model.}
\label{fig:Diffusefour}
\end{figure}
}

\newcommand{\placefigDiffusefive}{
\begin{figure*}
\begin{center}
\includegraphics[height=\textwidth,angle=90]{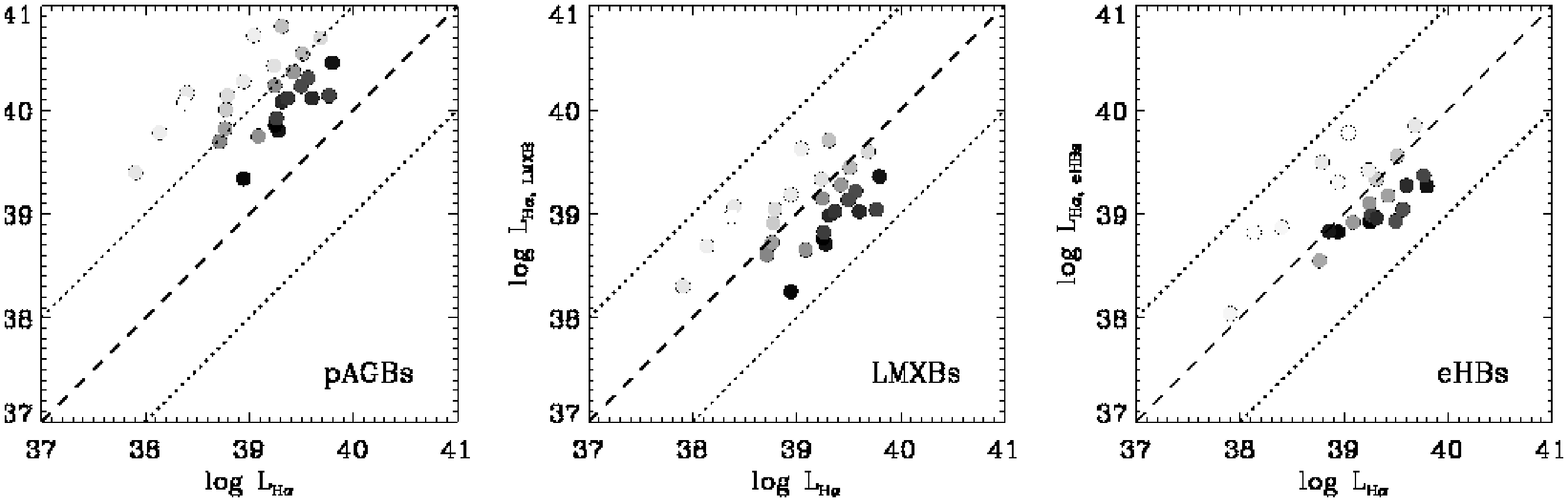}
\end{center}
\caption[]{
Ionisation balance for post-AGB stars, low-mass X-ray binaries and
extreme horizontal branch stars, within the region containing 90\% of
the nebular emission of our sample galaxies. Darker shades of grey
indicate galaxies with an higher area coverage of the ionised-gas,
where the ionising flux from the stellar sources should be
re-proccessed more efficiently. Objects where O- and B-stars have been
previously estiblished (\S\ref{subsec:Ostars}) as the main source of
ionisation (NGC~3032, NGC~4459, NGC~4526) are excluded from these
diagrams.}
\label{fig:Diffusefive}
\end{figure*}
}

\newcommand{\placefigShockone}{
\begin{figure}
\begin{center}
\includegraphics[width=\columnwidth]{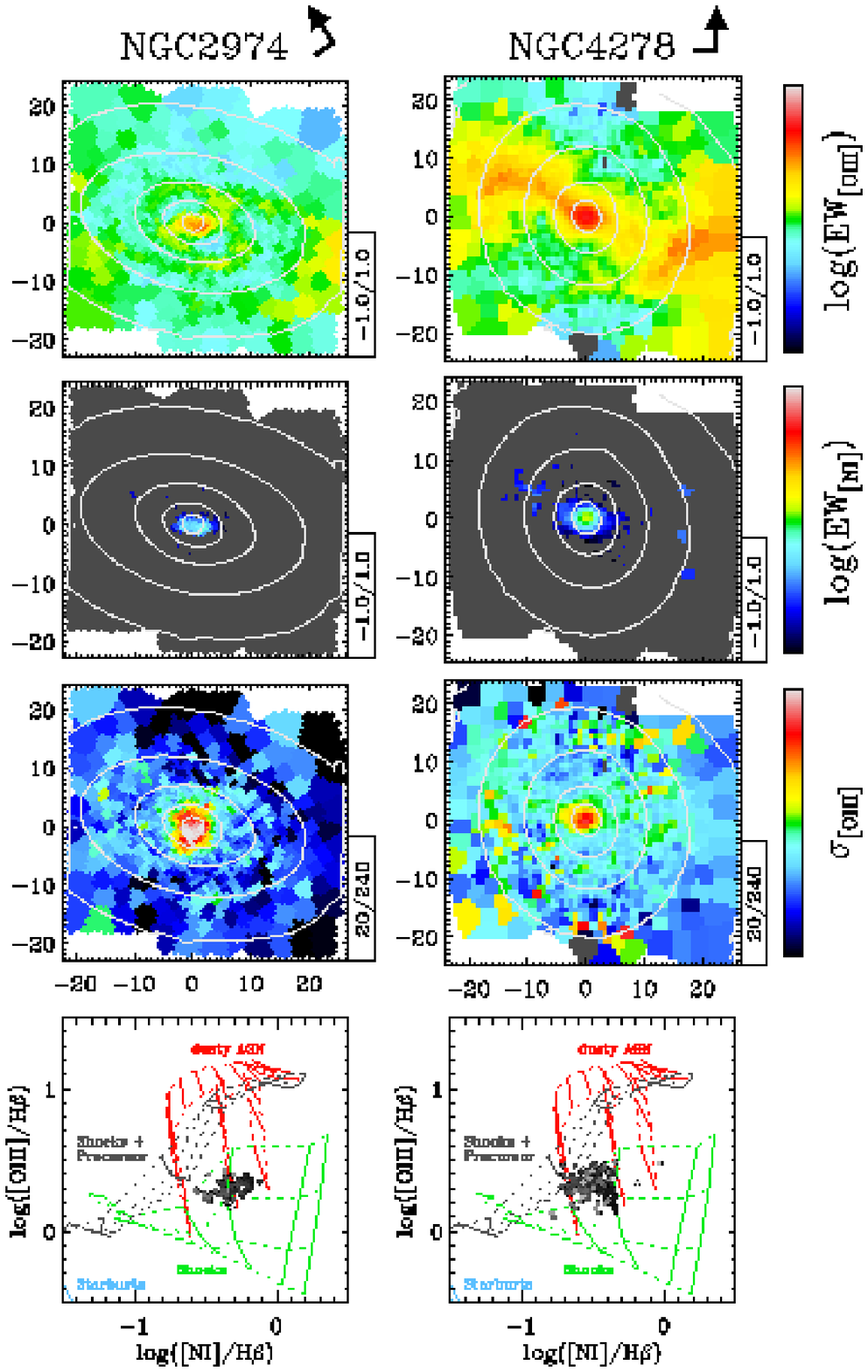}
\end{center}
\caption[]{
Two examples of elliptical galaxies where the presence of spiral-arms,
ovals, or integral-sign morphologies for the ionised-gas distribution
highlights the impact of non-axisymmetric perturbation of the
gravitational potential on the gas motions, possibly leading to shocks
between gas clouds.
Emission from the \Ni\ doublet 
(second row of panels)
is detected where the \Oiii\ (upper panels) and
\Hb\ emission is the strongest, allowing for a limited emission-line
diagnostic analysis (lower panels). The colored grids show the same
MAPPINGS-III models as in Fig.\ref{fig:NIone}, whereas the grey grids
display also the predictions for shock models with a gaseous precursor
ahead of the shock front.
These last grids are for models with the same values of the shock
velocity $V_s$ and the magnetic parameter $b$ as for the shock models
without precursor \Hii\ regions (see the caption of
Fig.~\ref{fig:NIone}).
The \NioHb\ diagnostic indicates that where the \Ni\ lines are
detected in NGC~2974 and NGC~4278 the observed emission is consistent
with both AGN photoinisation and shock excitation, although in the
latter case the required values for $V_s$ would range between 300 and
500 \kms, which can hardly be attained through gravitational motions
(see text).
Note that increasing the electron density of the schock models
($n_e=1$) would only shift the grid to the left as the \Ni\ lines
collisionally de-exites more easily, which would imply even higher
$V_s$ values to power the observed emission with shocks.
Maps for the gas velocity dispersion \Sg\ (third row of panels) are
also shown to illustrate how equivalent width structures (top)
are not always accompanied by obvious \Sg\ features, as expected if \Sg\
would trace $V_s$.
}
\label{fig:Shockone}
\end{figure}
}

\newcommand{\placefigHIERsone}{
\begin{figure}
\begin{center}
\includegraphics[width=\columnwidth]{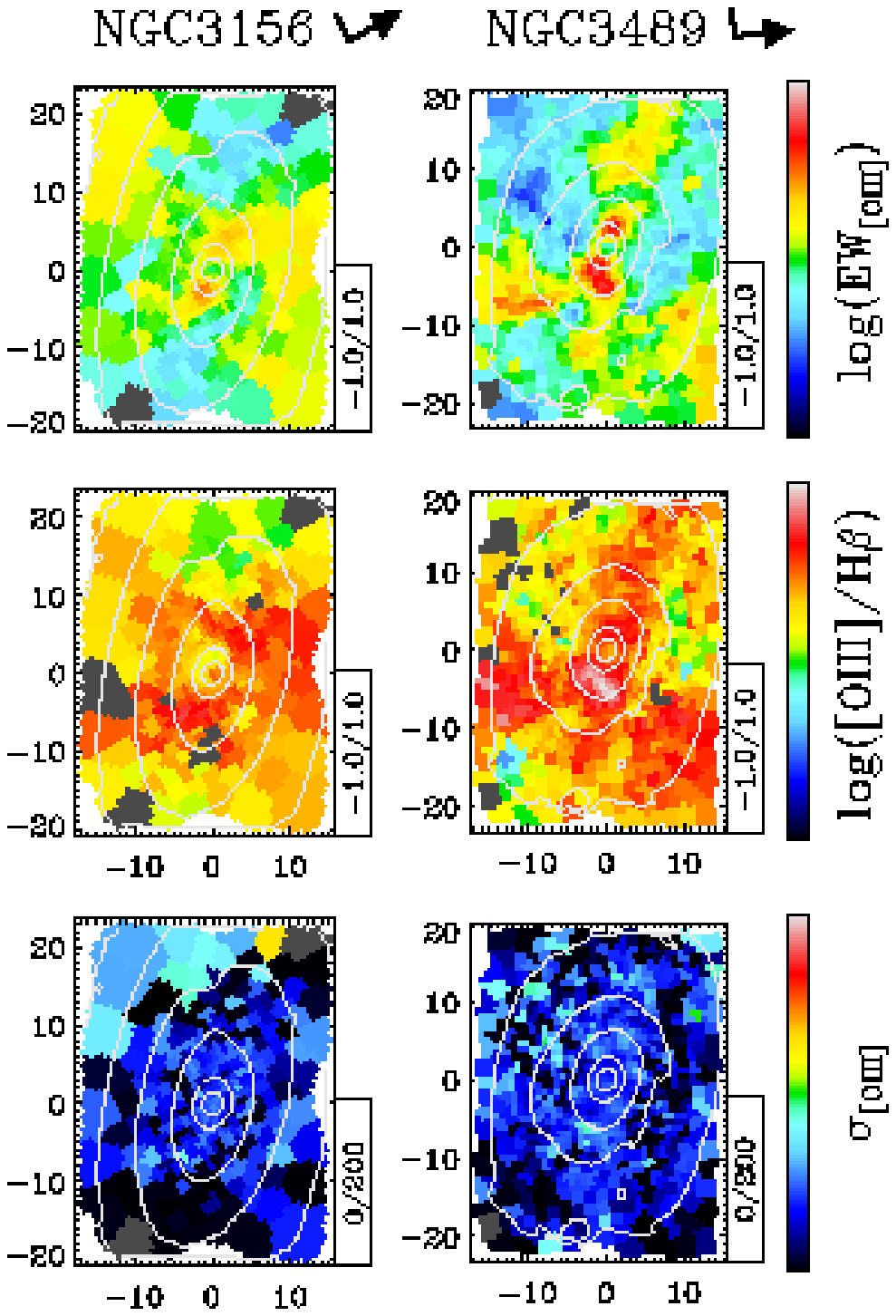}
\end{center}
\caption[]{
Two prototypical examples of early-type galaxies with intense nebular
emission characterised by large fluctuations for the \OiiioHb\ ratio,
which also peaks to relatively high values throughout extended
regions. From top to bottom, the panels show maps for the equivalent
width of the \Oiii\ line, for the \OiiioHb\ ratio and for the gas
velocity dispersion, respectively. Objects displaying such
highly-ionised extended regions (HIERs) are not very massive and show
invariably evidence for a recent star-formation episode. Both the
intensity of the emission and the large values for the \OiiioHb\ ratio
observed in HIERs can be attained considering as ionising sources the
pAGB stars associated to the recently formed stellar population (see
\S\ref{subsec:HIERs}), whereas the large fluctuations in the observed
\OiiioHb\ values is explained considering that star formation may
still be occurring in regions with very low \OiiioHb. The relaxed
character of the gas kinematics, as traced by the extremely small
values of the gas velocity dispersion, suggests that the conditions
for star formation are still favourable in the specific case of these
two objects.}
\label{fig:HIERsone}
\end{figure}
}

\newcommand{\placefigXrayone}{
\begin{figure}
\begin{center}
\includegraphics[width=\columnwidth]{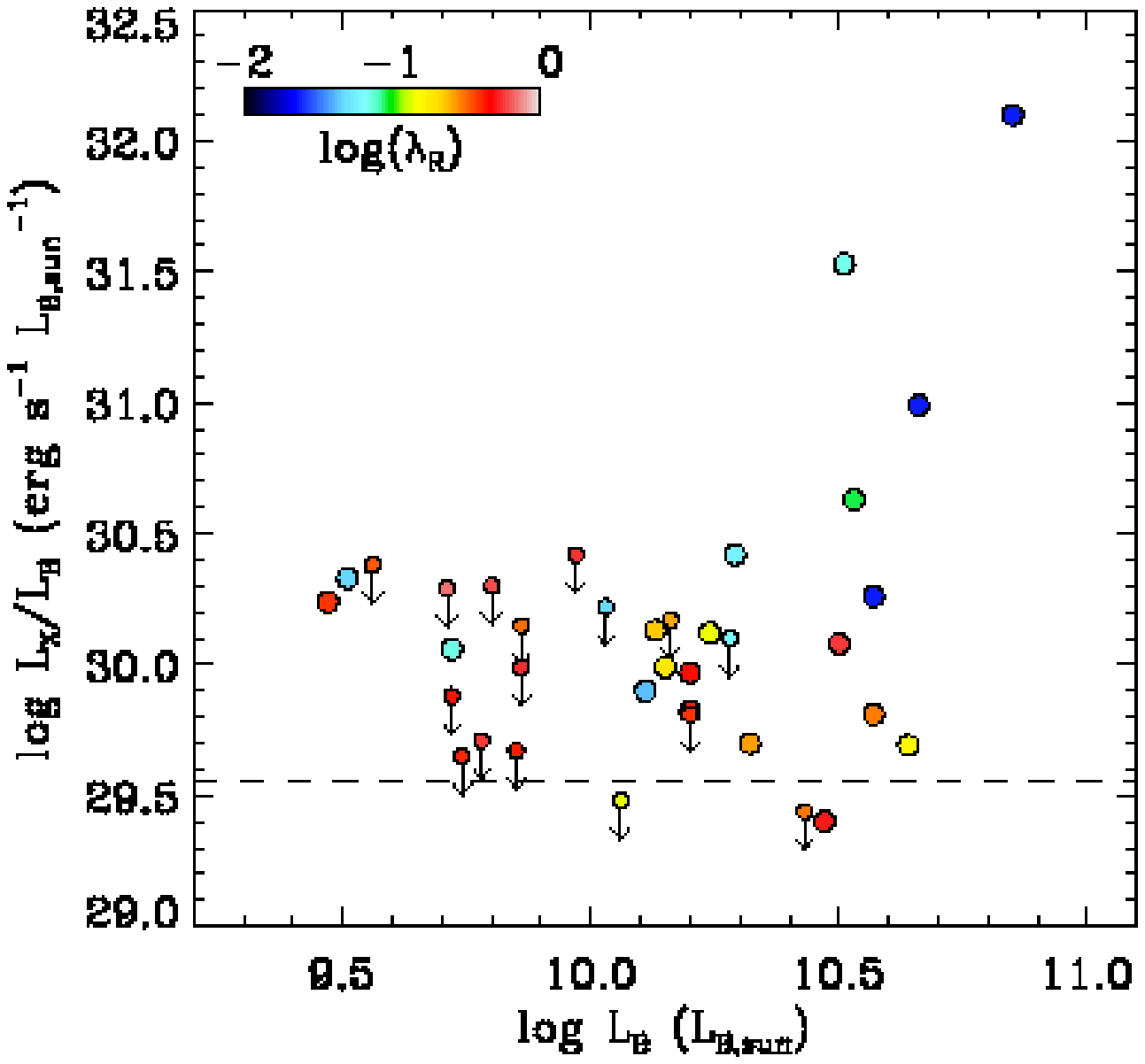}
\end{center}
\caption[]{
X-ray properties of \sauron\ early-type galaxies.
The total blue-band luminosity $L_B$ is compared to the X-ray
luminosity $L_X$, normalised to $L_B$. Symbols are colour-coded
according to the observed degree of rotational support as traced by
the $\lambda_R$ parameter of \citet{Ems07}, with green or blue symbols
showing slowly- or non-rotating objects with $\lambda_R < 0.1$.
$L_X$ values (circles) or upper limits (small circles with downward
arrows) are {\tt ROSAT\/} and {\tt Einstein\/} measurements from the
compilation of \citet{oSu01}, except for NGC524 and NGC5813, which
were taken from \citet{Pel05} and \citet{Boh00}, respectively.
These data cover three-quarters of the \sauron\ early-type sample,
with the remaining objects being typically low-luminosity galaxies
with an average $\log{\rm L_B}\sim$ 9.9.
The horizontal dashed line shows the normalised X-ray luminosity
expected from stellar sources \citep{oSu01}.
Only luminous galaxies can retain massive X-ray halos, in particular
if they are slowly- or non-rotating objects.}
\label{fig:Xrayone}
\end{figure}
}

\newcommand{\placefigXraytwo}{
\begin{figure*}
\begin{center}
\includegraphics[width=\textwidth]{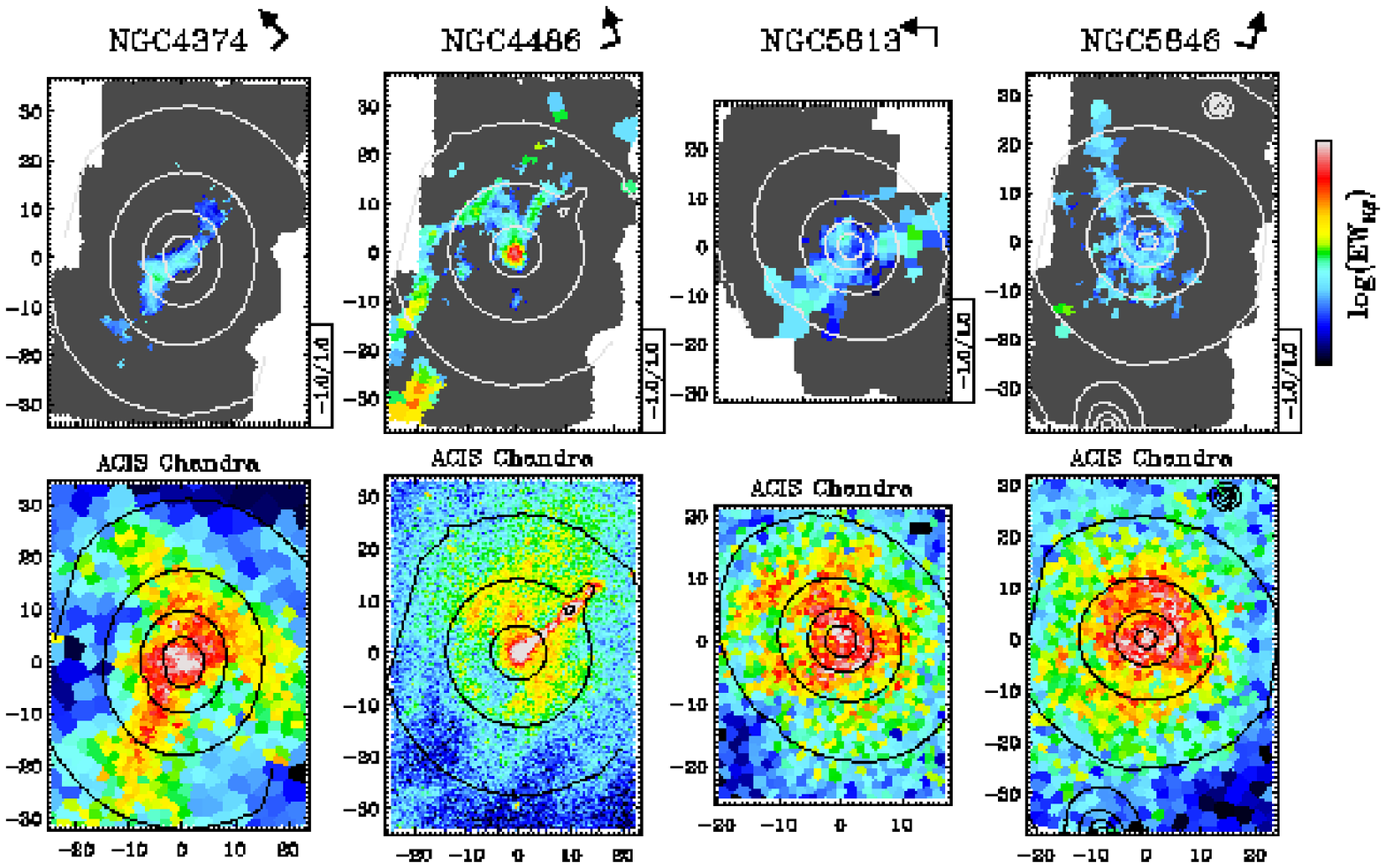}
\end{center}
\caption[]{
Warm {\it versus\/} hot gas emission in the four brightest and slowly-
or non-rotating galaxies in the \sauron\ galaxies, which all show also
prominent dust lanes.
The top panels show \sauron\ maps for the equivalent width of the
\Hb\ line, in \AA\ and on a logarithmic scale, whereas the lower
panels show Voronoi-binned maps for the X-ray emission observed with
Chandra, in the 0.3--5.0 keV band. Except for only a few regions, the
nebular emission is generally associated to X-ray emitting
features. For NGC~4486 and NGC~5846, this connection was already
discussed by \citet{Spa04} and \citet{Tri02}, respectively.}
\label{fig:Xraytwo}
\end{figure*}
}

\newcommand{\placefigXraythree}{
\begin{figure}
\includegraphics[width=\columnwidth]{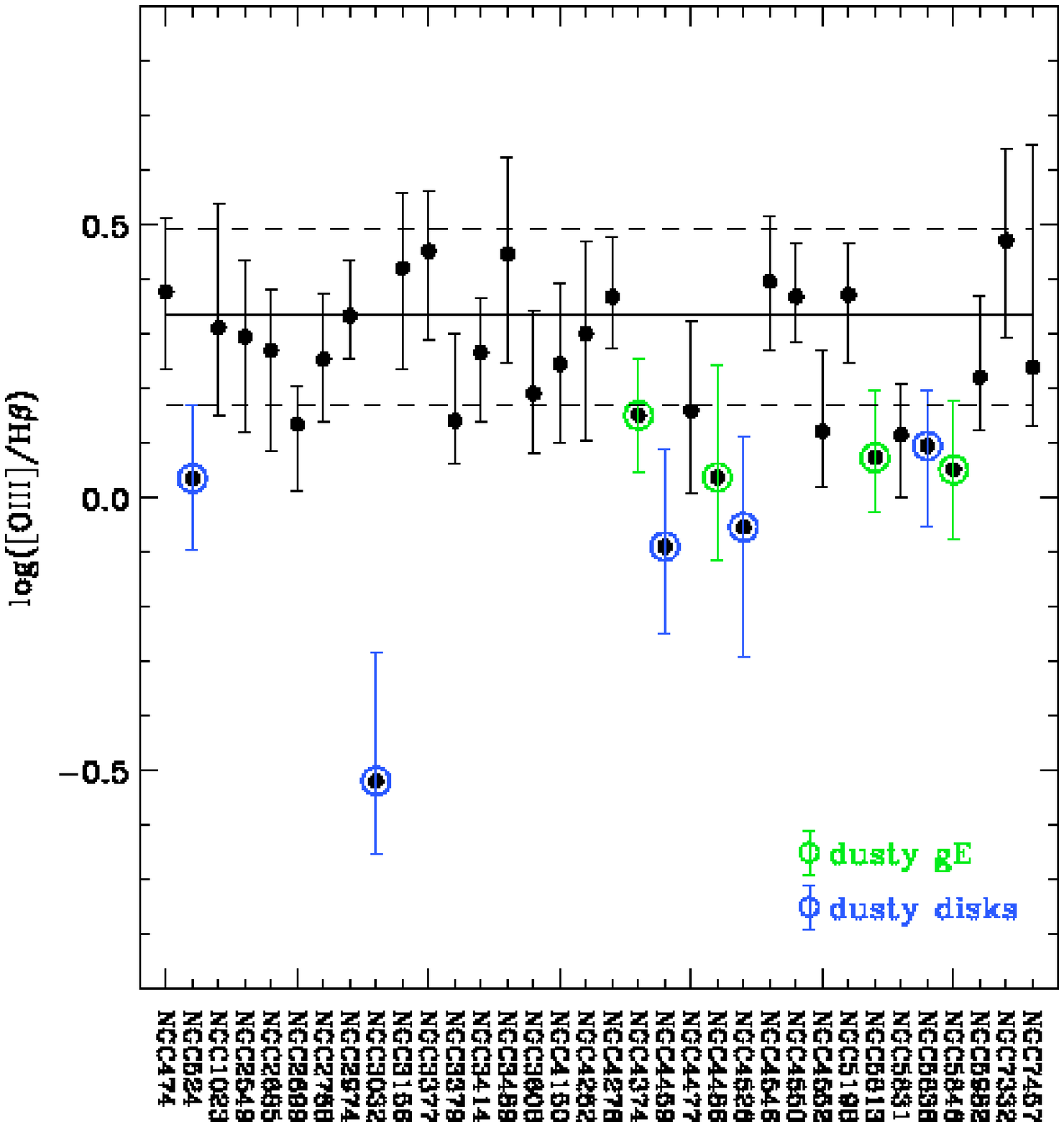}
\caption[]{
Median values and 1$\sigma$ confidence limits for the \OiiioHb\ ratio
observed in the \sauron\ galaxies with clearly detected and extended
\Hb\ and \Oiii\ emission, plotted in alphabetical order against the
galaxy name.  The brightest, dusty and slowly- or non-rotating
ellipticals are shown with green symbols, objects with dusty disks and
on-going star formation are shown in blue, and the remainder of this
subsample is denoted in black.  The horizontal dotted and dashed lines
show the average and 1$\sigma$ boundary for the \OiiioHb\ ratios
observed in the galaxies shown by the black points. The tendency for
the dusty giant ellipticals to show lower values for \OiiioHb\ ratio
than for the rest of the quiescent galaxies in the \sauron\ sample
suggests, in light of the presence in these objects of a massive halo
of X-ray emitting gas, that the interaction between the hot and warm
phases of the ISM has an impact on the ionisation of the nebular
plasma.}
\label{fig:Xraythree}
\end{figure}
}

\newcommand{\placefigSDSSone}{
\begin{figure}
\begin{center}
\includegraphics[width=\columnwidth]{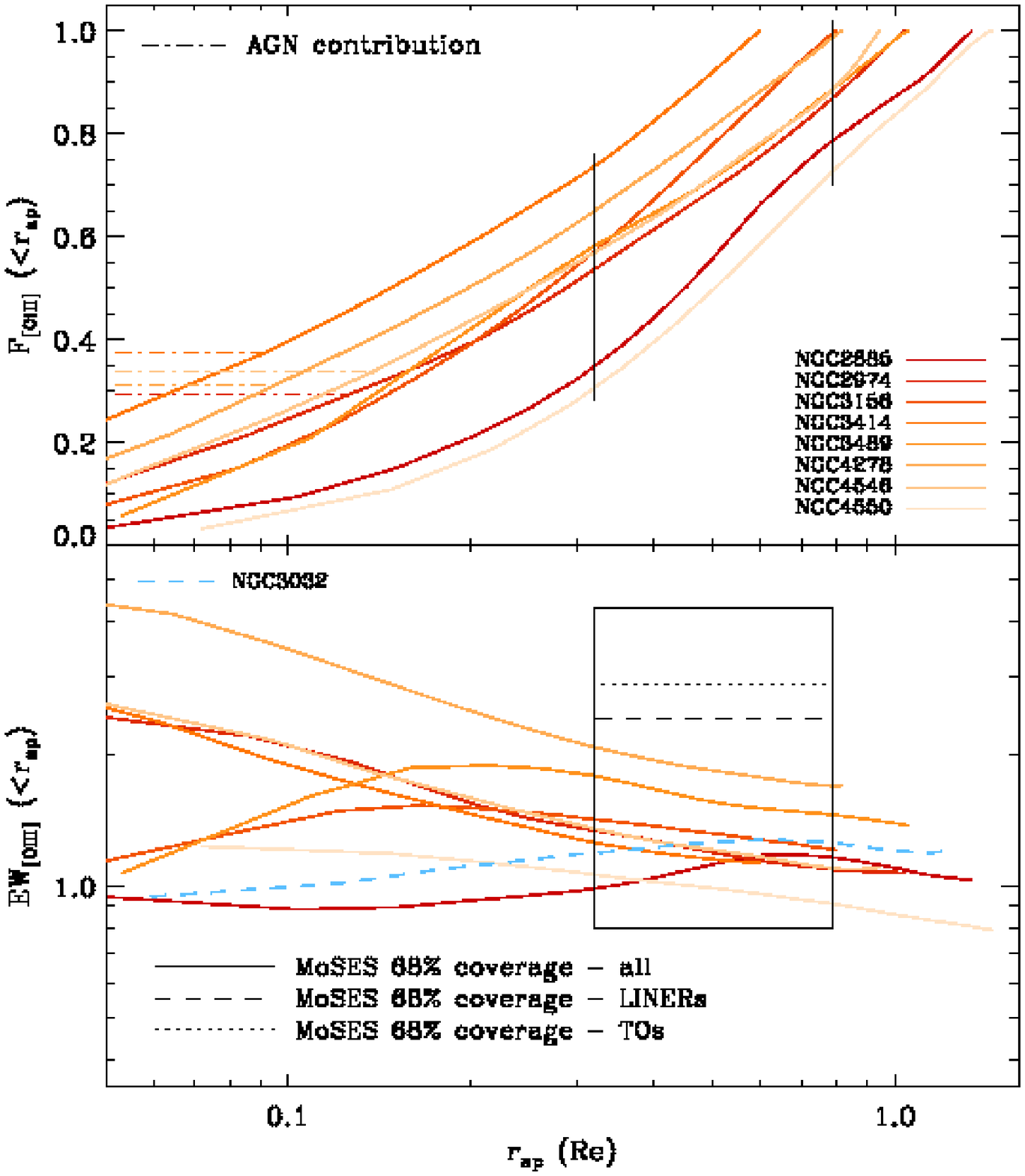}
\end{center}
\caption[]{
Radial profiles for the integrated flux (normalised to the total, top
panel) and equivalent width of the \Oiii~$\lambda5007$ emission (lower
panel) for the \sauron\ galaxies that display sufficiently intense
emission to be detected at the typical sensitivity of the SDSS spectra
for the MoSES early-type galaxy sample of \citet{Sch07}.
The integrated flux and equivalent width values are obtained over
circular apertures with radii expressed in units of the galaxy
effective radius $R_e$.
In the lower panel, the rectangle drawn with solid lines indicates
where 68\% of the values for the physical size of the galactic regions
that are encompassed by the $3''$-wide aperture of the SDSS spectra
are located, while also showing where 68\% of the values for the
equivalent width of the \Oiii\ emission detected in the MoSES spectra
can be found. In the same rectangle, the dotted and dashed horizontal
lines show the upper extent of the region containing 68\% of the
values for the equivalent witdh of the \Oiii\ emission observed only
in the MoSES galaxies displaying composite and LINER-like emission,
respectively.
In the upper panel, the dot-dashed horizontal lines show, for the
galaxies with Radio or X-ray signposts of nuclear activity, the most
conservative contribution of the AGN to the total \Oiii\ flux (that
is, within $3''$; see \S\ref{subsec:AGN}), whereas the vertical lines
show again the typical size of the galactic regions observed by the
MoSES spectra, which can be used to estimate the AGN contribution to
the nebular emission measured in such SDSS data.}
\label{fig:SDSSone}
\end{figure}
}